\documentclass[aps, pra, twocolumn, letterpaper]{revtex4}
\usepackage{graphicx}
\usepackage{indentfirst}
\usepackage{float}
\usepackage[top=2.5cm, left=2.0cm, right=2.0cm, bottom=2.5cm]{geometry}

\usepackage{pdflscape}
\usepackage{amssymb}
\usepackage{array}
\usepackage{mathtools}
\usepackage{dsfont}
\usepackage{amsfonts}
\usepackage{cancel}
\usepackage{bm}
\usepackage{color}
\usepackage{todonotes}
\usepackage{graphicx}

\newcommand{\reffig}[1]{Figure \ref{#1}}
\newcommand{\reffigP}[2]{Figure \ref{#1}(#2)}
\newcommand{\req}[1]{Equation (#1)}

\begin{document}

\title{Shaping Polaritons to Reshape Selection Rules}
\author{Francisco Machado$^{*1,2}$, Nicholas Rivera$^{*2}$, Hrvoje Buljan$^3$, Marin Solja\v{c}i\'{c}$^2$, Ido Kaminer$^{2,4}$}
\affiliation{
  $^{1}$Department of Physics, University of California, Berkeley, CA 97420, USA\\
  $^{2}$Department of Physics, Massachusetts Institute of Technology, Cambridge, MA 02139, USA  \\
  $^{3}$Department of Physics, University of Zagreb, Zagreb 10000, Croatia.\\
  $^{4}$Department of Electrical Engineering, Technion, Israel Institute of Technology, Haifa 32000, Israel.}
\clearpage

\begin{abstract}
  The discovery of orbital angular momentum (OAM) in light established a new degree of freedom by which to control not only its flow but also its interaction with matter. Here, we show that by shaping extremely sub-wavelength polariton modes, for example by imbuing  plasmon and phonon polariton with OAM, we engineer which transitions are allowed or forbidden in electronic systems such as atoms, molecules, and artificial atoms. Crucial to the feasibility of these engineered selection rules is the access to conventionally forbidden transitions afforded by sub-wavelength polaritons. We also find that the position of the absorbing atom provides a surprisingly rich parameter for controlling which absorption processes dominate over others. Additional tunability can be achieved by altering the polaritonic properties of the substrate, for example by tuning the carrier density in graphene, potentially enabling electronic control over selection rules. Our findings are best suited to OAM-carrying polaritonic modes that can be created in graphene, monolayer conductors, thin metallic films, and thin films of polar dielectrics such as boron nitride. By building on these findings we foresee the complete engineering of spectroscopic selection rules through the many degrees of freedom in the shape of optical fields.
\end{abstract}
\maketitle

\section{Introduction}
\label{sec-1}

The discovery that light can possess orbital angular momentum (OAM) \cite{Allen1992} beyond its intrinsic spin value of $\hbar$  has brought forth a new degree of freedom for the photon. By engineering the shape of optical modes, a wide variety of new applications has been developed including angular manipulation of objects \cite{He1995}, angular velocity measurement \cite{Lavery2013}, higher bandwidth communication using novel multiplexing techniques \cite{Yan2014, Wang2012} (already with on-chip implementations \cite{Ren2016}), quantum information systems \cite{Molina-Terriza2007}, quantum memory \cite{Nicolas2014} and sources of entangled light \cite{Mair2001}, which are beneficial to quantum cryptography implementations \cite{Mirhosseini2015, Malik2016}.

These results open the doors for another important application of shaped optical modes, tailoring the interactions between electrons and photons by enhancing or supressing electronic transitions. For example. when imbuing an optical mode with OAM  one expects novel selection rules based on conservation of angular momentum. This would provide a rich new degree of freedom for spectroscopy and many other studies where optical excitation is relied on for studying and controlling matter. Unfortunately, such  control over electronic transitions is not expected to be experimentally accessible because "the effective cross section of the atom is extremely small [compared to the wavelength of light]; so the helical phase front [of an OAM-carrying beam of light] is locally indistinguishable from an inclined plane wave" \cite{Yao2011}. In other words, the length scale mismatch between the electronic and photonic modes leads to a very small interaction beyond what can be observed in most experiments. Such a prediction has been corroborated in several theoretical studies \cite{Picon2010, Picon2010a, Afanasev2013,afanasev2016high} and is now taken as a basic fact \cite{asenjo2014dichroism}. As a result no proposals have been put forth for realistically controlling a wide variety of electronic transitions using the OAM of the photon.

\begin{figure}
  \centering
  \includegraphics[width = 0.4\textwidth]{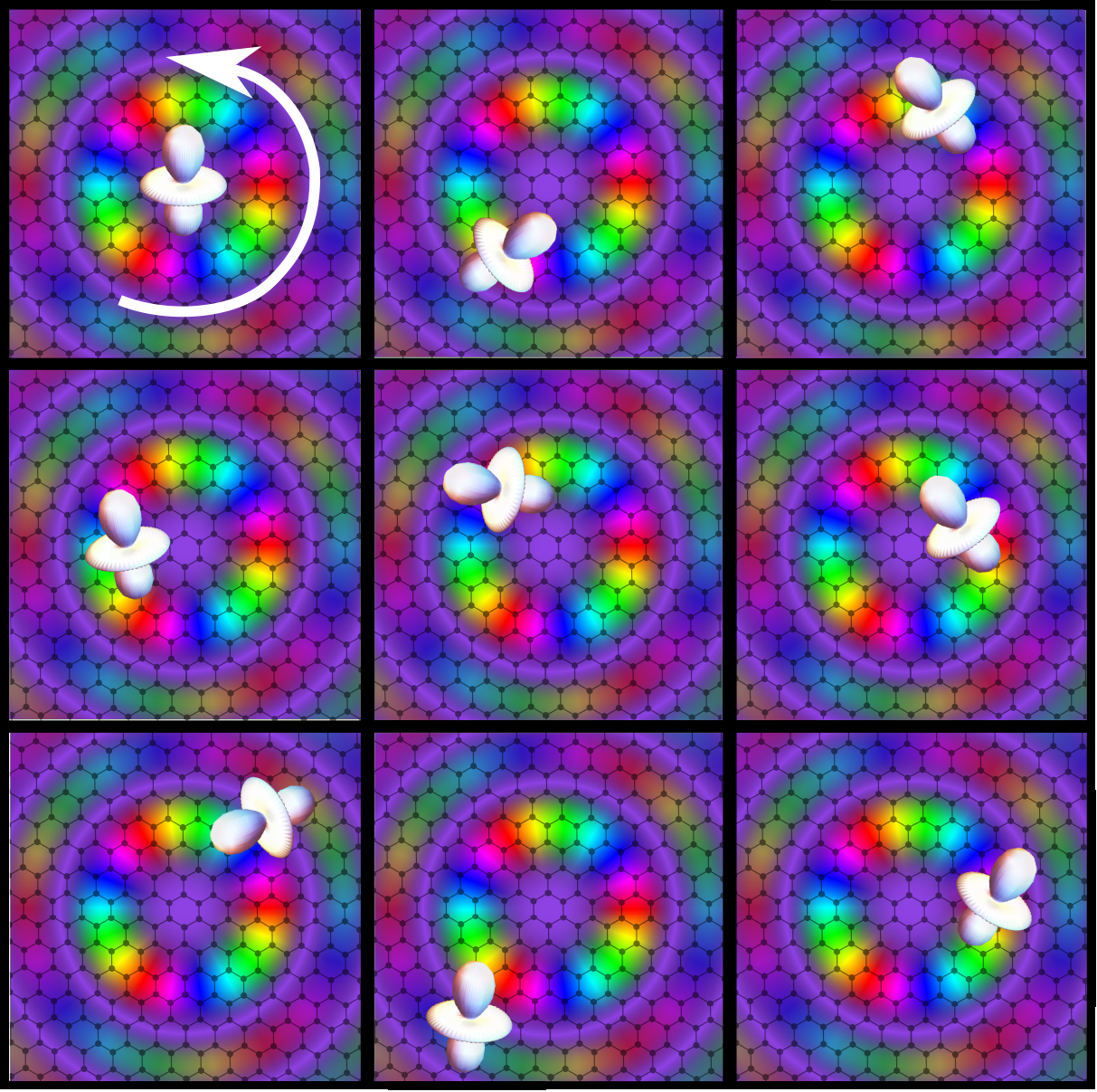}
\caption{{\bf Illustration of a polaritonic vortex and the basic setup for its interaction with an atomic system.} An electronic system, such as an atom (in white), is placed near the center of an OAM carrying polaritonic vortex mode (pictured with OAM = $3\hbar$), exciting transitions via selection rules based on conservation of angular momentum. By generating many parallel OAM carrying polaritonic modes one can enhance the resulting signal.}
  \label{fig:Schematic}
\end{figure}

The reasoning and results above are not only applicable to OAM-carrying light, but also any form of optical field shaping, thus implying the general infeasibility of tailoring electronic selection rules with shaped optical fields. That being said, there can exist exceptional cases where quadrupole transitions can be interfaced with OAM~\cite{Schmiegelow2016}. However these methods are very challenging experimentally and do not extend to other electronic transitions. Therefore, new approaches are necessary to use shaped light to control and explore forbidden electronic transitions in a wide variety of atoms, molecules, and other electronic systems. In such an approach, it is absolutely necessary to bridge the disparate length scales of the electronic system and photon.

Recent discoveries in polariton physics, ranging from surface plasmon polaritons (SPPs) in graphene \cite{Jablan2009, Ju2011, Chen2012, Grigorenko2012}, to surface phonon polaritons (SPhPs) in polar dielectric films \cite{Caldwell2015, Hillenbrand2002, Greffet2002}, may help bridge this length scale discrepancy as a result of their extremely short wavelengths, predicted to potentially be as small as just a few nanometers. For example, plasmon and phonon polaritons found in materials such as graphene \cite{Liu2008,Jablan2009,Fei2011, Fei2012, Ju2011, Chen2012, Grigorenko2012}, monolayer silver \cite{Nagao2001}, beryllium \cite{Diaconescu2007}, silicon carbide \cite{Caldwell2013}, and hexagonal boron nitride (hBN) \cite{Caldwell2014, Woessner2015, Tomadin2015} can have wavelengths 100-350 times shorter than the wavelength of a far-field photon at the same frequency. Several recent studies have predicted that as a result of these very short wavelength polaritons, not only could dipolar and quadrupolar transitions be greatly enhanced \cite{takase2013selection, Andersen2011, zurita2002multipolar,zurita2002multipolar2, filter2012controlling, jain2012near, rukhlenko2009spontaneous, yannopapas2015giant, Koppens2011}, but even high-order transitions that have never been considered could potentially become observable \cite{Rivera2016}. Despite these studies, no previous work has considered using the shape of these highly confined modes in order to efficiently enhance and suppress a wide variety of electronic transitions in a selective fashion.


In this paper we show that by shaping the polaritonic modes of a system we raise the possibility of tailoring the selection rules for absorption, using OAM carrying modes as an example. We show that these modes enable new selection rules based on the conservation of angular momentum, providing a new scheme for the efficient control of electronic transitions in atomic systems. We can use OAM-carrying polaritons to allow conventionally forbidden transitions to be fast and dominant and we can use OAM-carrying polaritons to forbid conventionally allowed transitions. We further find that tuning the placement of the absorber and the dispersion of the polaritons  provides a means to study many different electronic processes in a controllable way. The scheme we propose in this work may open the doors to  controlling electronic selection rules with the immense number of degrees of freedom in general shaped optical fields. In the long run, combining this technique with ultra-fast pulse technology can provide precise spatial-temporal control over the electronic degrees of freedom in myriad systems.


\section{Results}

\label{sec-3}

We start our analysis by considering the simplest scenario: an atom placed inside an OAM-carrying polariton mode such that the atom is concentric with the mode's center. This is illustrated schematically in the top left of \reffig{fig:Schematic}. For concreteness we consider the atomic system to be hydrogen. Although we use hydrogen as a particular example, the physics we demonstrate in this work can be readily extended to many other atomic and molecular systems, particularly those with spherical or axial symmetry. The OAM-carrying polariton mode, typically called vortex or vortex mode, can be constructed from the superposition of incoming plane wave polaritonic modes whose phase difference is proportional to the incoming angle as:
\begin{align}
  \bm{E}_{q,m}(\bm{\rho},z) &= E_0\int_0^{2\pi} \frac{d\alpha }{2\pi} \  e^{i(q\rho\cos(\phi-\alpha) - \omega t)} \frac{\hat{q} + i \hat{z}}{\sqrt{2}} e^{-qz} e^{i\alpha m}\notag\\
  &\Rightarrow  E_z \propto J_{m}(q\rho) e^{im\phi} e^{-q z} \quad (z>0),
\end{align}
where $\bm{E}_{q,m}(\bm{\rho},z)$ is the electric field profile of the mode, $J_m$ is the Bessel function of order $m$, $\bm{\rho} =\{\rho,\phi\}$ are the in-plane distance and angle, $z$ and $\hat{z}$ are the out-of-plane position and direction respectively, $\alpha$ is the angle of the incoming plane wave polariton in direction $\hat{q}$, $q$ its wavevector, $\omega$ its angular frequency, and $E_0$ its amplitude.

The most important parameters in our analysis are the confinement factor of the vortex and its OAM. The confinement factor $\eta$ measures how small the wavelength of the mode is relative to the wavelength of a free space photon of the same frequency, and is defined by $\eta = qc/\omega$.  The OAM of the mode is related to the phase winding of the vortex, $m$, by $\hbar m$. For example, in \reffig{fig:Schematic}, we show the phase profile of vortex modes with an OAM of 3$\hbar$. Throughout the text, we assume that the polariton vortex is created in a 2D plasmonic material with Drude dispersion. We arrive at quantitatively similar results for finite thickness substrates and other dispersion relations, such as those of phonon polaritons, provided that the confinement factors are the same. This is explained in detail in the Supplementary Materials (SM).

\begin{figure*}
 \centering 
  \includegraphics[width = 0.8\textwidth]{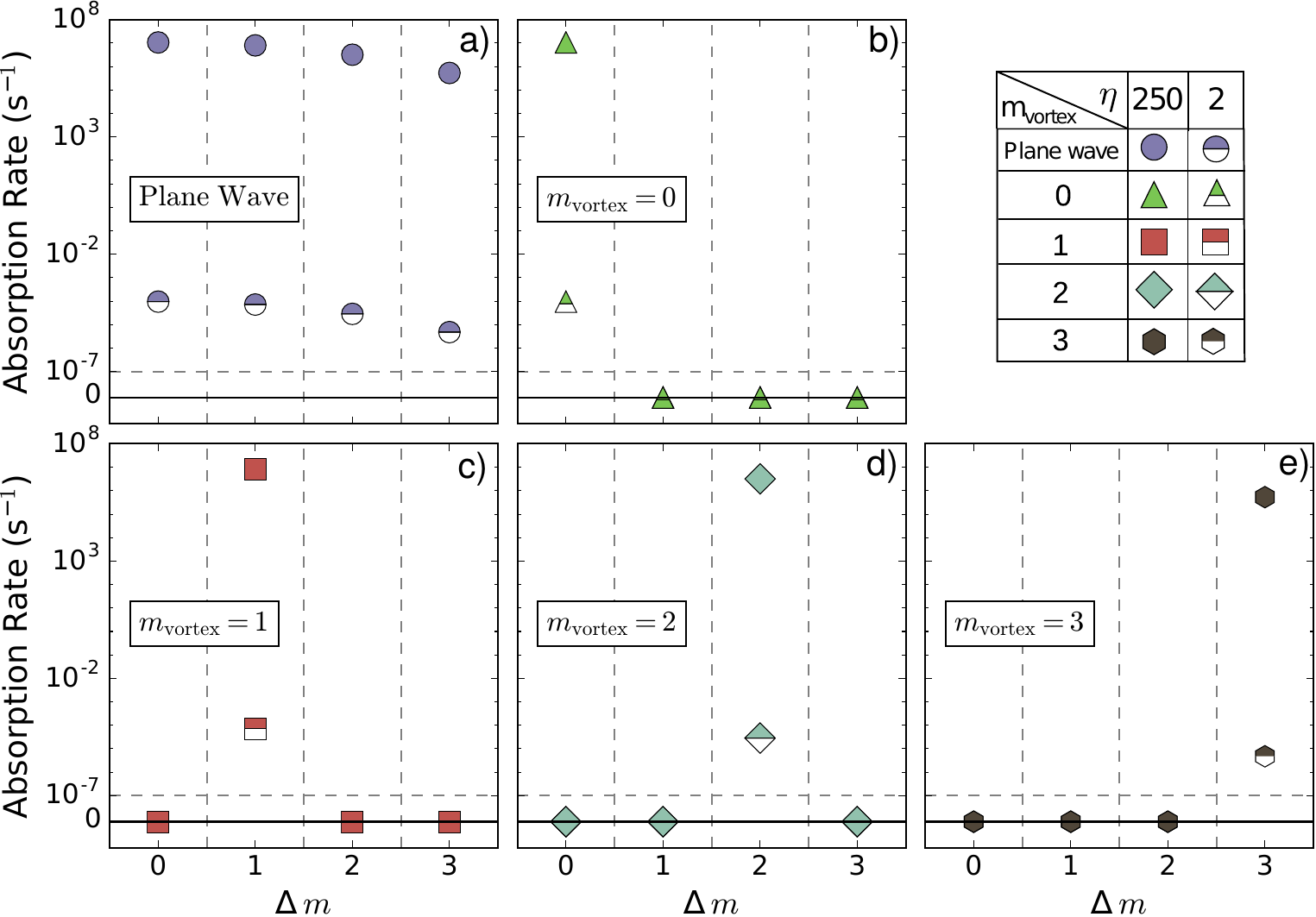}
  \caption{{\bf Selection Rules in the absorption of OAM carrying SP(h)P modes.} Calculation of the absorption rate due to a plane wave SP(h)P (a) and an OAM carrying vortex SP(h)P (b-e) for different transitions in the family $(5,0,0) \to (6,3,\Delta m)$ for two different values of confinement factor $\eta$, 2 (half-filled) and 250 (filled), with $z = 20$ nm. The vortex modes impose selection rules on the electronic transitions, while the increase in confinement factor leads to an enhancement of the absorption rate by a factor of  $\sim 10^{11}$. The examples of $\eta=2$ show that, although free space OAM carrying modes could in principle impose the same selection rules, the difference in length scales between the polariton and the atom results in absorption rates too small for experimental observation. The absorption rates are normalized by assuming each SP(h)P mode carries a single photon.}
    \label{fig2}
\end{figure*}

In \reffig{fig2}, we consider the effect of the vortex polariton modes on the absorption rates of different transitions, calculated to first order in perturbation theory. For concreteness, we consider a set of transitions between principal quantum numbers 5 and 6 (with a transition free-space wavelength of $7.45\  \mu$m). In the example explored in this figure, we consider the octupole (E3) transition between the $5s$ state and the $6f_{0,1,2,3}$ states. Normally such a transition is considered highly forbidden. The atom is taken to be $20$ nm away from the surface of the polariton-sustaining material. In \reffigP{fig2}{a}, we consider the absorption rate of a plane wave polaritonic mode and observe that transitions rates between different $\Delta m$ states are all non-zero. All transitions are allowed and there is no control over the electronic dynamics. In contrast, when considering the absorption of a vortex mode with $ m_{\text{vortex}} = 0,1,2,3$ in Figures~\ref{fig2}(b-e), respectively, only the transition corresponding to conservation of angular momentum is non-zero, the remaining transitions have been suppressed.

These selection rules arise naturally in the Fermi's Golden Rule formalism in a cylindrically-symmetric system. In this case the transition rate is proportional to the square of $\int d\phi\ e^{-i(m_f-m_i) \phi} e^{im_{\text{vortex}} \phi}$, where $\hbar m_i$ and $\hbar m_f$ are the $z$-projected angular momentum of the initial and final electron states, and $m_{\text{vortex}}$ is the phase-winding index of the polariton vortex mode. This proportionality makes it clear that the transition rate is zero unless:
\begin{equation}\label{eq:Sel}
\Delta m = m_f - m_i = m_{\text{vortex}}.
\end{equation}  
This simple equation tells us that differences in the $z$-projected OAM of the electron must be supplied by the vortex, thus justifying the interpretation of the phase winding as an angular momentum. The reason that plane waves do not yield the same level of control is that a plane wave is equivalent to a superposition of vortices with all possible angular momenta. Therefore, the angular momentum selection rule is always satisfied for some transition where an electron absorbs a plane wave. 

Although \req{2} tells us which transitions are allowed and which are not, it does not tell us in advance if the allowed transitions happen quickly enough to be observable. To this end, in \reffig{fig2}, we also quantify the rates of different electronic transitions due to the absorption of unshaped (plane wave) polaritons, as in \reffigP{fig2}{a}, and shaped (vortex) polaritons, as in Figures~\ref{fig2}(b-e). In both cases, the absorption rates are a sharp function of confinement. For a confinement factor of $2$, the non-zero E3 transitions rates are on the order of 1 event per 10 hours while at a confinement of $250$, the rates are on the order of 1 event per 300 nanoseconds (for an atom 20 nm away from the material's surface), an 11 order of magnitude enhancement. As the atom gets even closer to the surface (for example when the atom is 5 nm away), the rate can increase to nearly 1 event per 500 picoseconds, which would be considered fast even for (free-space) dipole transitions. These quantitative results make it evident that angular momentum conservation alone is insufficient to ensure experimentally accessible selection rule modification: it is also imperative to match the scale of the atom and photon.

A notable consequence of the results presented in \reffig{fig2} is that using SP(h)P vortex modes yields \emph{highly enhanced and controllable} electronic transitions for arbitrarily large values of $\Delta m$, in contrast to the case with plane wave (unshaped) polariton modes. In particular, by creating a vortex with a fixed OAM and placing atoms concentric with that vortex, it is possible to forbid conventionally allowed dipole transitions (by making the OAM of the vortex greater than $1$), and allow conventionally forbidden multipole transitions, thus providing a way to access and control conventionally disallowed transitions which are normally invisible in spectroscopy.

Before proceeding to further analyze the ability to control electronic transitions with shaped optical modes, we pause to discuss some experimental considerations in implementing the above results. Schematically, an experiment to observe the effects described in this paper would feature: a polariton-sustaining surface, absorbing atoms, and a means to create vortex modes with different angular momenta in the vicinity of these atoms. The goal is to observe the modified absorption of the sample as a function of the angular momentum of the vortex, which can be indirectly probed by monitoring the fluorescence of the sample as a result of its interaction with the vortex. An important requirement of the polariton-sustaining surface is that its wavelength is comparable to the size of the orbitals of the atoms. Because of that, optimal materials for creating vortices include ultra-thin films of plasmonic (gold, silver), phononic (silicon carbide, boron nitride) materials, or 2D conductors like graphene. Creating the vortices can be done by illuminating an appropriately shaped grating coupler near the surface, such as those used in Refs.\cite{David2015, Spektor2015, David2016, Spektor2017}. Of course, the radius of the coupler should be comparable to the polariton wavelength so that the polariton does not decay completely when propagating to the center. We assume this situation throughout the text. Finally, we note that while polaritonic vortices have been demonstrated in plasmonic materials such as gold and silver, they have yet to be demonstrated in graphene or phonon-polariton materials. Therefore, a first and exciting step towards implementing our proposed scheme is to generate vortices in these materials. 

Regarding the choice of absorbing atom, because the results of \reffig{fig2} are for an atom centered with the vortex, our scheme is most cleanly implemented on atom-like systems whose placement can be controlled very well, such as quantum dots. Another potential advantage of quantum dots is that they can be made to have mesoscopic sizes, allowing one to match the length scales of the vortex with the size of the atom-like system more easily (i.e., a smaller polariton confinement is required). For example, for a practicable quantum dot size of 20 nm \cite{andersen2011strongly}, one should be able to access and control conventionally forbidden transitions through the use of vortices of $70$ nm wavelength, which have already been demonstrated on thin silver films \cite{David2015}. For even larger quantum dot sizes, it is conceivable that the same effects we describe here could be observed by using vortices of conventional surface plasmons even on thick metallic films.

However, while the scheme we propose can be more readily implemented with artificial atoms, it would be of appreciable interest to implement it with the much smaller natural atomic and molecular systems, whose myriad forbidden multipolar transitions have been elusive to spectroscopy since its early days. And while it is potentially possible to find polaritons that are sufficiently highly confined to be interfaced with atomic and molecular emitters \cite{Rivera2016}, a challenge in implementing our scheme arises from the considerable uncertainty in the placement of the atom. Therefore, we now study the effect of off-center displacement of the atom on the prospects for access and control over forbidden transitions.

When an atom is off-center from the vortex, the rotational symmetry around the vortex center is broken, meaning that angular momentum conservation no longer holds. More explicitly, the absorption rate of a vortex by an off-center atom is no longer related to $\int d\phi\;e^{-i(m_f-m_i)\phi}e^{im_{\text{vortex}}\phi}$ but rather $\int d\phi e^{-i(m_f-m_i)\phi} \sum\limits_{m = -\infty}^{\infty} C_{m-m_{\text{vortex}}}^{\phi_0} e^{im\phi}J_{m-m_{\text{vortex}}}(qD) $, where $q$ is the wavevector of the polariton vortex mode, $D$ is the radial separation between the atomic and the vortex center and $C_{\delta m}^{\phi_0}$ is a complex number of unit magnitude ($|C_{\delta m}^{\phi_0}|=1$) dependent on $\delta m = m-m_{\text{vortex}}$ and the angular position of the atom $\phi_0$. The full derivation can be found in the SM. As a result, the rate of absorption of a single polariton vortex (of angular momentum $\hbar m_{\text{vortex}}$) at a distance $D$ for a transition between states $i$ and $f$ with a change in the $z$-projected angular momentum $\hbar\Delta m$, denoted $\Gamma_{m_{\text{vortex}}}(i\rightarrow f,D)$, becomes:
\begin{equation}\label{eq:disp}
\Gamma_{m_{\text{vortex}}}(i\rightarrow f,D) = J^2_{\Delta m-m_{\text{vortex}}}(qD)\Gamma_{\Delta m}(i\rightarrow f,0),
\end{equation}
where $\Gamma_{\Delta m}(i\rightarrow f,0)$ is the rate of absorption of a single vortex polariton when the atom is aligned with the vortex mode of OAM $\hbar \Delta m$. Note that for $D=0$ (centered atom), the rate is only non-zero when the transition satisfies $\Delta m=m_{\text{vortex}}$, in agreement with \req{2}. \req{3} is physically consistent with an analogous result that has been obtained in the context of the absorption of far-field twisted light by atomic systems \cite{Scholz-Marggraf2014}.

\begin{figure*}
  \centering
  \includegraphics[width = 0.7\textwidth]{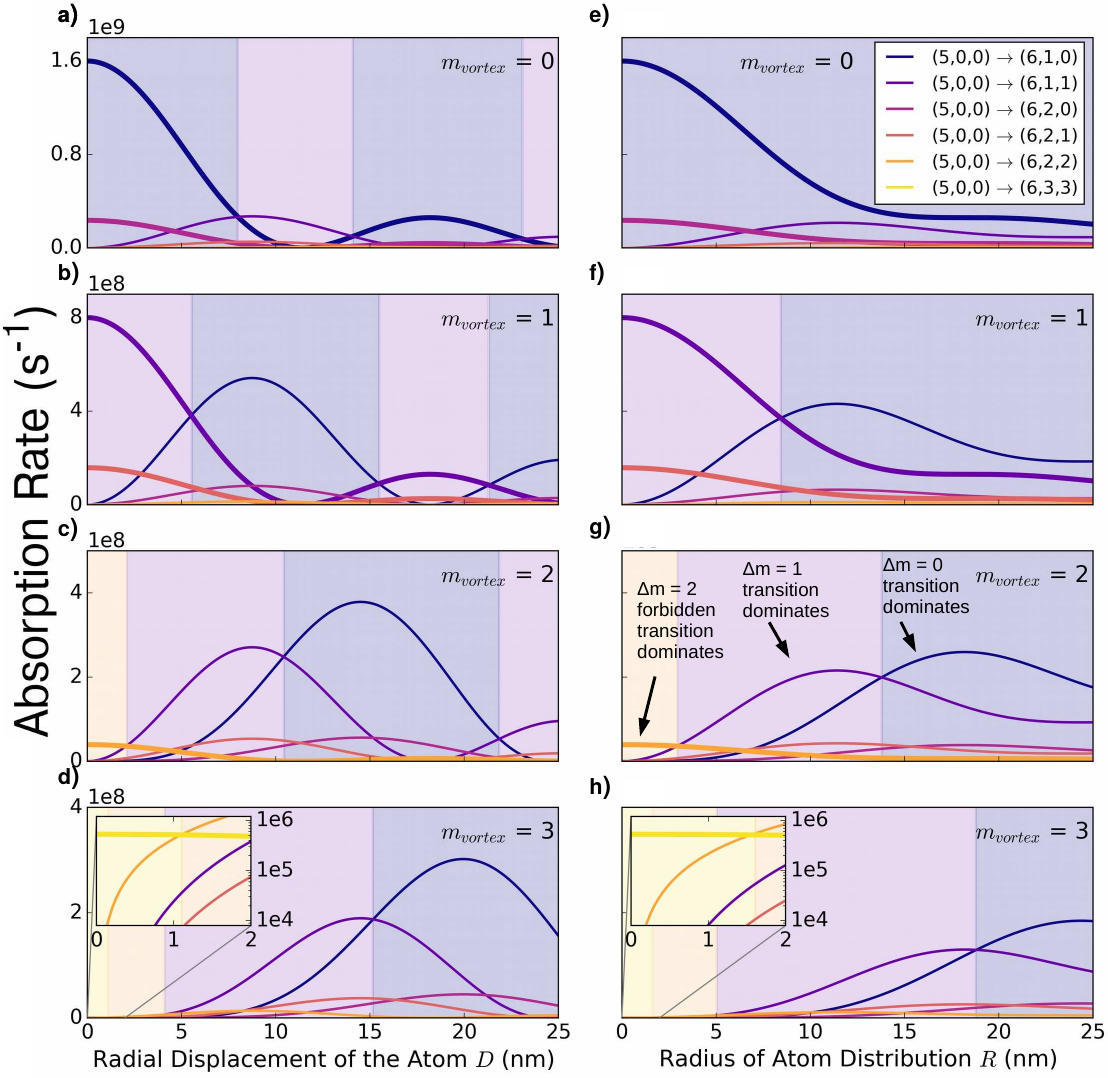}
  \caption{{\bf Robustness of the angular momentum  selection rules to atom displacement.} Dependence of the absorption rates of a displaced individual atom (left column) and uniform atomic distribution (right column) for transitions with initial state (5,0,0) and final principal quantum number 6 at a confinement factor of $\eta=250$ and $z = 20$ nm. As the rotational  symmetry is broken, the selection rule discussed in \reffig{fig2} is no longer valid and all $\Delta m$ transitions become allowed. The absorption rates match the selection rule of \req{2} at $D=0$, and for small $D$ this transition always dominates. For larger $D$, other transitions can become dominant, as the value of $J_0(qD)$ decreases and higher order Bessel terms become comparable. The background color of each plot corresponds to the dominant transition in that regime. The angular momentum-conserving transitions are highlighted by a thicker line.}
  \label{fig3}
\end{figure*}
\begin{figure}
  \centering
  \includegraphics[width = 0.5\textwidth]{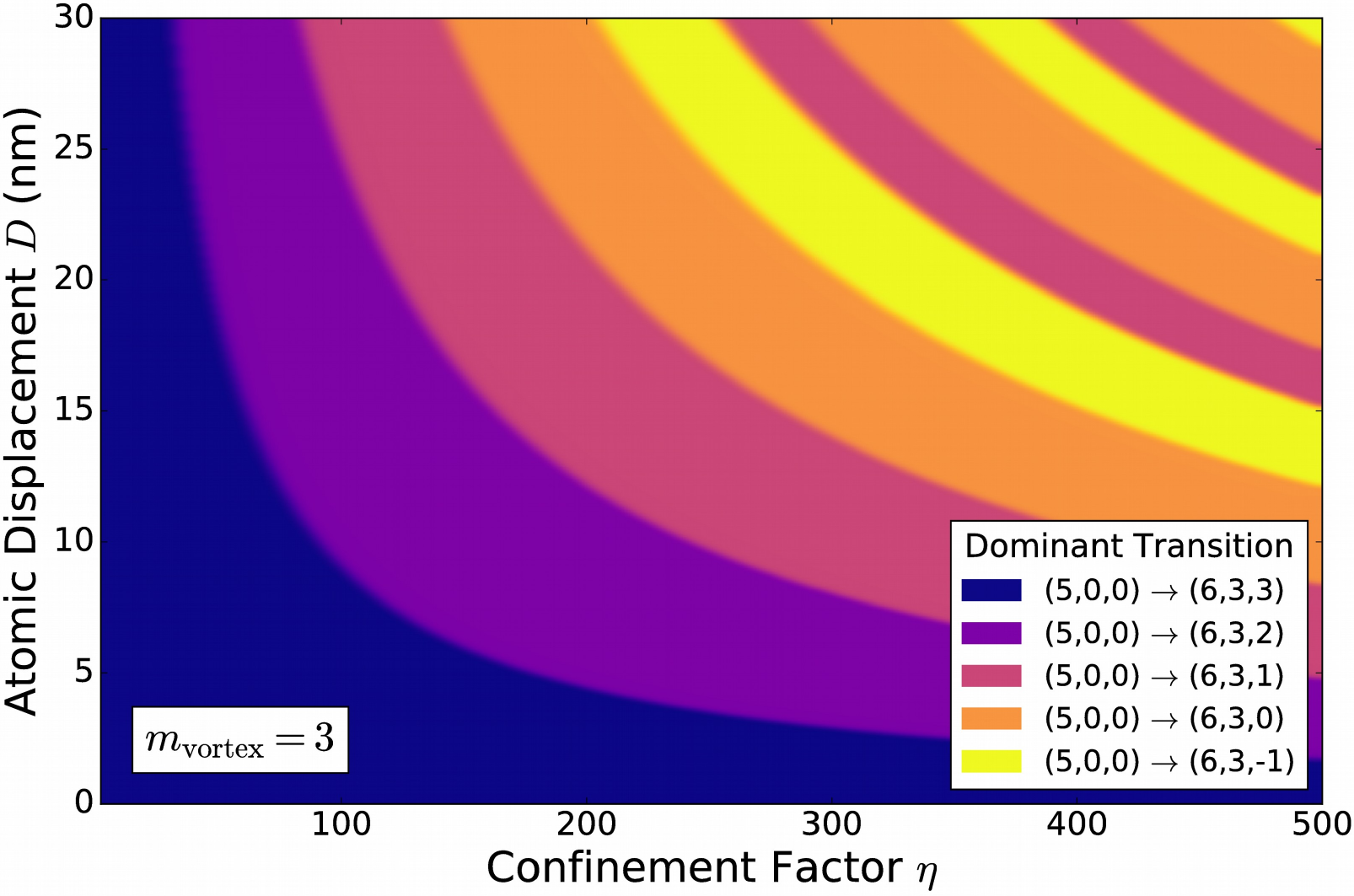}
  \caption{{\bf Landscape of dominant transitions for $m_{\text{vortex}} = 3$.} For a hydrogen atom at $z=20$ nm, the dominant process in the family $(5,0,0)\to(6,3,\Delta m)$ is plotted as a function of the confinement factor $\eta$ and the displacement of the atomic system $D$. For a given vortex mode, the displacement and confinement can be used to control the dynamics of the electronic system. This is of particular interest in materials such as graphene, where the confinement factor of the mode can be electrically tuned, opening the possibility for electrical control over atomic transitions.}
  \label{fig4}
\end{figure}

\req{3}, while a rather simple result, contains much interesting physics which Figures \ref{fig3} and \ref{fig4} intend to summarize. The most obvious consequence of \req{3} is that when the atom is off-center relative to the vortex  ($D\neq 0$), the non-zero value of the Bessel function, $J_{\Delta m-m_{\text{vortex}}}(qD)$, allows for transitions that were forbidden when $D=0$. Since the argument of the Bessel function in \req{3} is $qD$, the strength of the different absorption processes varies at the length scale of the wavelength of the vortex mode and not the size of the atom. This means that the interaction is robust to misalignments of nanometers or even tens of nanometers, and is not sensitive to fluctuations on the atomic scale. For the remainder of this section, we shall refer to the originally allowed transitions for $D=0$ as angular-momentum-conserving, and the originally disallowed ones as non-angular-momentum-conserving. 


In Figures \ref{fig3}(a-d), we calculate the rates of different absorption transitions between initial state 5s and final states with principal quantum number 6  as a function of atomic displacement $D$ for varying vortex OAM (increasing from top to bottom). For zero radial displacement, only angular-momentum-conserving transitions  (highlighted by a thicker line) have a non-zero absorption rate, as dictated by \req{2}. As the radial displacement of the atom is increased, the rate of non-angular-momentum-conserving transitions increases and eventually may dominate over angular-momentum-conserving transitions. Although this appears to reduce control over electronic transitions, because of the oscillatory behavior of the Bessel function with displacement, there are regimes where one particular transition (though not necessarily the angular-momentum-conserving one) dominates over all others - providing a parameter to tune which transition dominates. These regimes in Figures \ref{fig3}(a-d) are marked using the colored background, where the colors denote which transitions dominate. Additionally, a single vortex can be used to "turn off" certain absorption transitions because the absorption rate has nodes where the Bessel function vanishes, as can also be seen in Figures \ref{fig3}(a-d). This means that, somewhat surprisingly, a \textit{single vortex} can actually be used to controllably (through $D$) access and study different transitions, beyond what angular-momentum conservation dictates. We conclude the analysis of Figures \ref{fig3}(a-d) by pointing out that this controlled access is yet again only possible because the rates of transitions calculated in these figures are sufficiently high that these transitions can be observed. Were the confinement factor 2, the transitions would be 11 orders of magnitude slower, and this degree of control using atomic placement would be rendered inaccessible.



The analysis presented in Figures \ref{fig3}(a-d) is pertinent to the case when a single atom is placed exactly at a radial displacement $D$ from the center of the vortex, which is expected to well-characterize a mesoscopic absorber with a more precise placement, such as a quantum dot. In Figures \ref{fig3}(e-h), we consider the case where a population of randomly distributed absorbing atoms interacts with the polariton vortex. In these panels, the average rate of absorption of the sample is computed as a function of vortex angular momentum and the radius $R$ of the random uniform distribution of atoms, which is set by the deposition conditions. The effect of the distribution is to average the results of Figures 3(a-d) over $D$. Despite this averaging, $R$ can still be used as a reliable parameter for tuning the dominant transition in the system, as highlighted in the changing colored backgrounds in Figures 3(f-h). Therefore it still is the case that the radius of the distribution / sample size can be used to make non-angular-momentum-conserving transitions dominant, thus acting as a parameter to control the strengths of different once-forbidden transitions.

In \reffig{fig3}, we have discussed the effect of the radial displacement $D$, but \req{3} tells us that the wavevector of the vortex mode serves equally well as a tuning parameter by which to control which transitions dominate and which ones do not. In polaritonic platforms such as graphene, it is already possible to tune the wavevector of the plasmon at a fixed frequency by changing the carrier concentration (or equivalently Fermi energy $E_F$). The wavevector depends on the doping level in graphene via $q = \frac{2\alpha\omega}{\bar{\epsilon}_rc}\frac{\hbar\omega}{E_F}$\footnote{This formula is correct within the Drude model formalism with some correction typically at larger values of $q$.}, where $\bar{\epsilon}_r$ is the average dielectric constant surrounding the film \cite{Fei2012,Jablan2009} and $\alpha$ is the fine-structure constant. In \reffig{fig4}, we explore the consequences of tuning the polariton wavevector on controlling forbidden electronic transitions. We consider the absorption rate of a polariton vortex (for a fixed OAM of $3\hbar$) for different transitions ($5s \rightarrow 6f_{\Delta m}$) as a function of the radial displacement $D$ and the confinement factor $\eta=\frac{qc}{\omega}$ which directly modifies the wavevector $q$. The colored regions indicate which transition of this family dominates. As can be seen, for fixed radial displacement, as the confinement (or Fermi energy) is tuned, we can choose which transition dominates. In fact, at a radial displacement of 30 nm, we can switch between five different transitions by tuning the confinement factor between 1 and 250. Analogous to Figures \ref{fig3}(e-h), when considering a uniform distribution of absorbers, the radius of this distribution still provides a good parameter for controlling the dominant transition, as illustrated in the SM. More detailed information regarding the rates of these different transitions can also be found in the SM.


  
Before concluding we note that in order to tune the confinement factor while keeping the same vortex angular momentum, one would need a near-field coupling scheme such as a circular grating where the phase of the emission at different points in the grating is set so that the phase winds $m_{\text{vortex}}$ times, irrespectively of the confinement factor. This ensures that the relative phases of the in-coupled (plane wave) polaritons are fixed as we vary the confinement factor of the mode, thus retaining the necessary interference to build the vortex. One way to potentially  achieve this is by illuminating a circular grating with OAM-carrying far-field photons so that the angular momentum of the light (up to the polarization angular momentum) is imprinted onto the grating.

\section{Discussion}
In summary, we have shown that polaritons with angular momentum allow for access to and control over absorption processes as a result of both conservation of angular momentum and extreme sub-wavelength confinement. This holds most readily when the absorbing atom is aligned with the vortex center. Nevertheless, control over electronic transitions (including non-angular-momentum-conserving ones) can even be obtained when atoms are randomly distributed in the vortex through two tuning parameters: polariton dispersion and distribution size. The conclusions we arrive at hold for a wide range of polaritonic materials, whether they be plasmon, phonon, exciton, or other classes of polaritons.

Finally, we discuss interesting effects of light-matter interactions with vortex polaritons beyond single-photon processes. In the SM, we extend our formalism to consider the case where the atom absorbs two quanta of light, as would be the case in multiphoton spectroscopy. We find that when an atom (concentric with vortices) absorbs two vortex modes, the angular momentum selection rule becomes $m_f - m_i = m_{\text{vortex},1}+m_{\text{vortex},2}$, where $\hbar(m_f-m_i)$ is the change in $z$-projected orbital angular momentum of the electron and $\hbar m_{\text{vortex},1}$, $\hbar m_{\text{vortex},2}$ are the angular momenta of the two absorbed vortices. Therefore, it should be possible to tailor the selection rules of multiphoton spectroscopy through the use of two vortices with potentially different angular momenta. We find that the strength of the multiphoton interaction is strong for highly confined vortex modes.


In addition to absorption, we also considered spontaneous emission of vortex modes by an atom in order to pave the way for doing near-field quantum optics with vortex modes. We find that for emitters interfaced with polaritonic materials sustaining highly confined modes, the spontaneous emission into one vortex (with high OAM) and into two vortices (with OAM $0$ or $\pm \hbar$) can be quite fast. However, one would need an emitter with a highly $m$-dependent energy spectrum such that the decay rate of the emitter depends sharply on the OAM of the emitted polariton. In such a circumstance, one could have an emitter which selectively generates a single vortex quantum with a desired value of OAM or even potentially entangled vortex quanta.

In the long term, the ability to engineer the electronic transitions in a quantum system, enabled by polaritonic modes, opens the doors for many applications which depend on usually inaccessible quantum states. Generating these quantum states in simple table-top settings will lead to novel light emitting devices and even lasing technologies, by enabling new decay paths in quantum systems. At the same time, including OAM carrying polariton modes in the toolbox of spectroscopy adds a new technique with which to probe and investigate electronic transitions and states, in particular in multi-electron systems, where the large degeneracy of the states is lifted and an even greater control over the electronic transitions is allowed. In another direction, exploring stronger fields in OAM carrying polaritons will give us access to regimes of non-perturbative physics, where the electronic states themselves are being altered. More generally, we believe this technique holds promising prospects for using the complete set of degrees of freedom in the temporal and spatial shaping of optical fields, for coherent control and engineering of the electron dynamics in quantum systems.

\section{Methods}

We analyze the light-matter interaction by writing down the electromagnetic field operator in the basis of vortex modes and then using the resulting interaction Hamiltonian to compute the rates of various interaction processes between light and matter using Fermi's Golden Rule. The Hamiltonian and corresponding field operators are given below as:
\begin{align} 
H &= H_{\text{ele}} + H_{\text{SP(h)P}} + H_{\text{int}}\quad \text{ with} \notag\\
  H_{int} &= \frac{e}{2m_e} (\bm{A}\cdot \bm{p} + \bm{p} \cdot \bm{A})+ \frac{e^2}{m_e^2}\bm{A}^2\quad\text{ where } \notag\\
  \bm{A} &= \sum_{q,m} \sqrt{\frac{\hbar q^2}{4\bar{\epsilon}_r \epsilon_0 \omega_q L\xi_q}} (\bm{F}_{q,m}\hat{a}_{q,m} + \bm{F}_{q,m}^* \hat{a}^\dag_{q,m} ) \label{eq:QuantizedField},
\end{align}

\noindent where $H_{\text{ele}}$ is the Hamiltonian of the electron, $H_{\text{SP(h)P}}$ is the Hamiltonian of the SP(h)P modes, and $H_{\text{int}}$ is the interaction Hamiltonian between the electron and SP(h)Ps. $m_e$ and $e$ are the mass and charge of the electron, $\epsilon_0$ the vacuum permittivity, $\bar{\epsilon}_r$ the average relative permittivity of the dielectric above and below the interface and $L$ the quantization length of the system.  $\bm{A}$ corresponds to the vector potential operator and it is written as an expansion over dimensionless modes of the vector potential $\bm{F}_{q,m}$, with corresponding creation (annihilation) operators $\hat{a}_{q,m}^\dag$ ($\hat{a}_{q,m}$). These modes can be derived from the integral expression in \req{1}:

\begin{widetext}
  \begin{equation}
    \bm{F}_{q,m}= e^{-q|z|}  \frac{1}{2\sqrt{2}} e^{im\phi} i^{m+1} \Bigg(  \hat{\rho} [J_{m+1}(q\rho) - J_{m-1}(q\rho)] -i \hat{\phi} \frac{2m J_{m}(q\rho)}{q\rho} +
    \left. \begin{cases}
      \hat{z}2J_{m}(q\rho) & (z>0)\\
      0 & (z=0)\\
      - \hat{z}2J_{m}(q\rho) & (z<0)
    \end{cases}\right),
  \end{equation}
\end{widetext}
where $J_m(q\rho)$ is the Bessel function of order $m$, $\rho, \phi, z$ are the standard cylindrical coordinates and $q$ is the wavevector of the mode. The parameter $\xi_q$ is a dimensionless normalization factor which is required for the energy of the vortex mode to be $\hbar\omega$. We find that the factor $\xi_q v_g$, where $v_g = d\omega/dq$ is the group velocity, dictates the strength of light-matter interactions and is similar for polaritons in different materials with the same confinement factor.  The details of our calculations are provided in the SM.



\begin{acknowledgments}
  This work was partially motivated by discussions with Moti Segev on whether the orbital angular momentum of light is an intrinsic property of the single photon or a joint property of the ensemble of photons.
  Research supported as part of the Army Research Office through the Institute for Soldier Nanotechnologies under contract no. W911NF-13-D-0001 (photon management for developing nuclear-TPV and fuel-TPV mm-scale-systems). Also supported as part of the S3TEC, an Energy Frontier Research Center funded by the US Department of Energy under grant no. DE-SC0001299 (for fundamental photon transport related to solar TPVs and solar-TEs).
  N.R. was supported by a Department of Energy fellowship no. DE-FG02-97ER25308.
  H.B. acknowledges support from the QuantiXLie Center of Excellence.
  I. K. was partially supported by the Seventh Framework Programme of the European Research Council (FP7-Marie Curie IOF) under Grant agreement No. 328853-MC-BSiCS.
\end{acknowledgments}


\bibliography{Main_text.bib}

\end{document}


\title{Supplementary Material for: \\ Shaping Polaritons to Reshape Selection Rules}
\author{Francisco Machado$^{*1}$, Nicholas Rivera$^{*1}$, Hrvoje Buljan$^2$, Marin Solja\v{c}i\'{c}$^1$, Ido Kaminer$^{1,3}$}
\affiliation{$^{1}$Department of Physics, Massachusetts Institute of Technology, Cambridge, MA 02139, USA  \\
  $^{2}$Department of Physics, University of Zagreb, Zagreb 10000, Croatia.\\
  $^{3}$Department of Electrical Engineering, Technion, Israel Institute of Technology, Haifa 32000, Israel.}

\maketitle

\section{Introduction to Supplementary Material}

In this Supplementary Material (SM) we first use a different family of transitions in hydrogen to explore the impact of vortex modes on the electronic selection rules of an atom-like system. We also demonstrate that the results presented in \rfig{4} of the main text also apply when we consider not only an absorber at a particular position but one with a distribution over its positions. These additional data further supplement the conclusions of our work.

The rest of the SM details the derivation of expressions for rates of transitions in which an atom absorbs (or emits) a polariton with orbital angular momentum. We will focus on the particular case of 2D plasmons described by a local Drude model and hyperbolic surface phonon polaritons in uniaxial hexagonal boron nitride (hBN). However, we present our results in a form that allows for straightforward extension not only to other conductivity models for 2D plasmonic materials but also to other surface polaritons such as surface exciton polaritons. We begin by constructing the classical vortex modes for polaritons that are highly confined. Then we develop quantized field operators in the basis of modes with definite orbital angular momentum and use these operators to compute transition rates at first and second order in perturbation theory.

\section{Appendix 1: Supplementary Data}

\subsection{Selection rules for transitions up to $\Delta m=4$}

\begin{figure}
  \centering 
  \includegraphics[trim={0 0 0 0.05cm},clip,width = 0.8\textwidth]{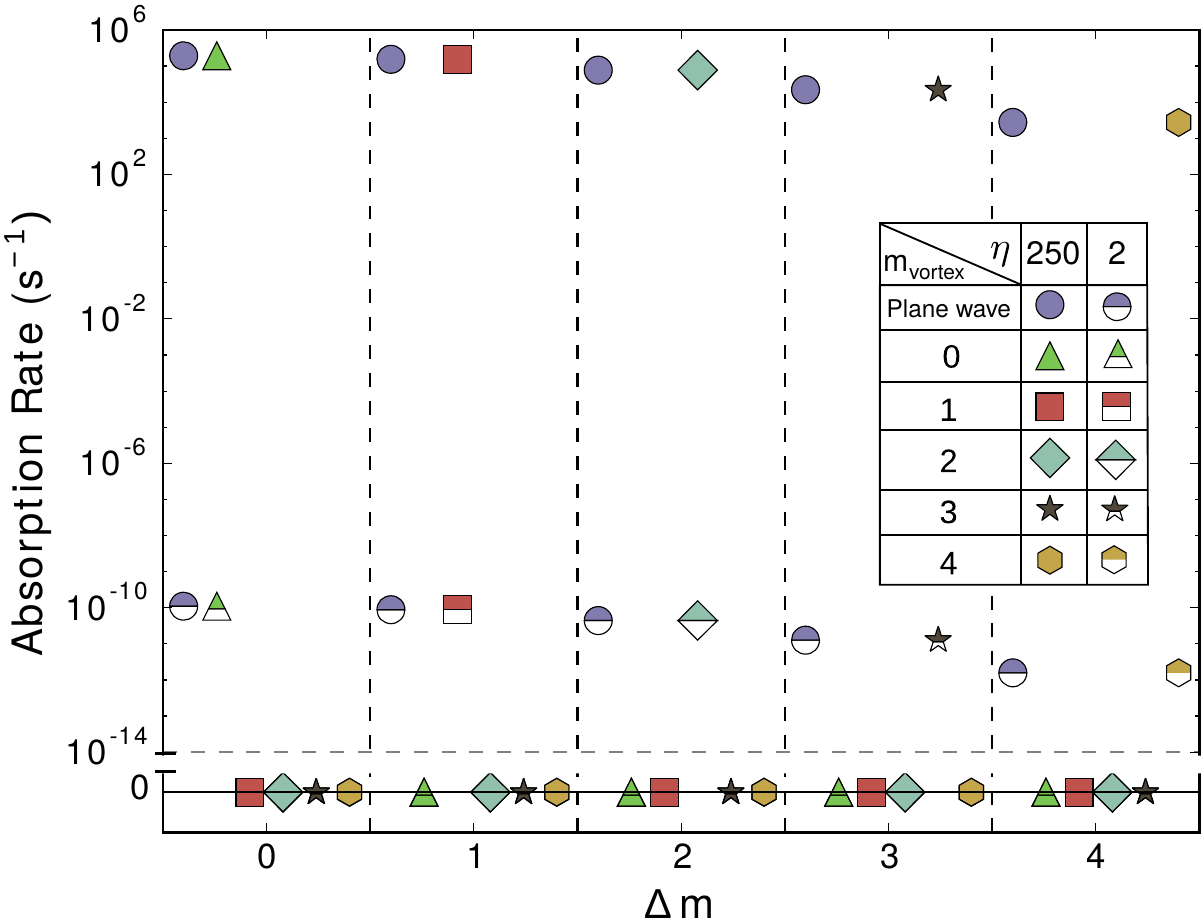}
\caption{{\bf Selection Rules in the absorption of OAM carrying SP(h)P modes.} Calculation of the absorption rate due to a planar SP(h)P and an OAM carrying vortex SP(h)P for different transition in the family $(5,0,0) \to (6,4,\Delta m)$ for two different values of confinement factor $\eta$, 2 (half-filled) and 250 (filled), with $z_0 = 20$ nm. The vortex modes impose selection rules on the electronic transitions, while the increase in confinement factor leads to an improvement of the absorption rate by a factor of  $\sim 10^{15}$. The examples of $\eta=2$ show that, although free space OAM carrying modes could in principle impose the same selection rules, the difference in length-scales between radiation and the atom size results in absorption rates too small for experimental observation. The absorption rates are normalized by assuming the SP(h)P modes carries a single photon quanta.}
  \label{fig2_Sup}
\end{figure}

In \rfig{2} of the main text we explored the selection rules of vortex modes in the family of transitions $(5,0,0)\to(6,3,\Delta m)$ of hydrogen. In this part of the SM we consider the family of transitions $(5,0,0)\to(6,4,\Delta m)$, as to demonstrate the generality of the before mentioned selection rules. In \reffig{fig2_Sup} we consider the transition rates between $(5,0,0)\to(6,4,\Delta m)$ due to the absorption of a plane wave polariton or a vortex mode polariton. As in \rfig{2}, the plane wave polariton enables all the transitions in this family (albeit with very different rates), while the vortex modes only enable transitions with $\Delta m = m_{vortex}$. Moreover, by considering transitions with a different $\Delta l$, we also emphasize the greater impact the confinement factor has on the absorption rate when compared with \rfig{2}. For concreteness, if we consider the dipole forbidden transition $(5,0,0)\to(6,4,0)$, (a E4/hexadecapole transition), the increase of the confinement factor from 2 to 250 leads to an increase in the transition rate from an event every many hundreds of years to one every 10\ $μ$s, a 15 order of magnitude enhancement. In the case of \rfig{2}, where $\Delta l = 3$, we observe an increase of 11 orders of magnitude.

\subsection{Absorption rates of $m_{\text{vortex}} = 5 $ SP(h)P mode with starting state (5,0,0)}
\begin{figure}
  \centering
  \includegraphics[width = 0.8\textwidth]{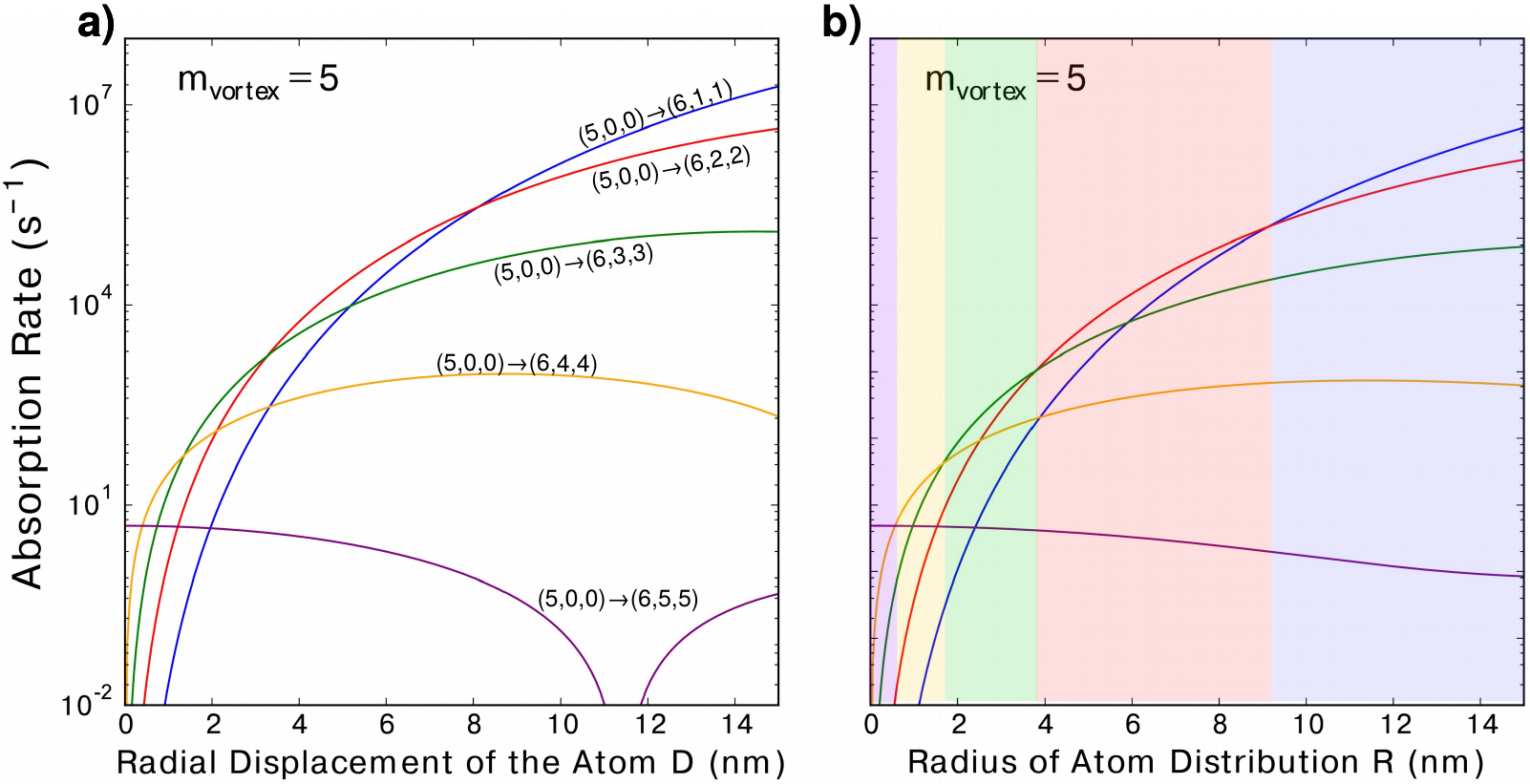}
  \caption{ {\bf Absorption rates for a highly suppressed transition.} Using the same parameters as in \rfig{3}, the absorption of $m_{\text{vortex}}=5$ SP(h)P modes is computed. The left column presents the absorption rates as a function of the atom displacement $D$ while the right column presents the rates as a function of the radius $R$ of an uniform distribution of the position of the atom. The colored regions represent the dominance of the differently colored transitions. }
  \label{fig4_old}
\end{figure}

\begin{figure}
  \centering
    \includegraphics[width =0.9\textwidth]{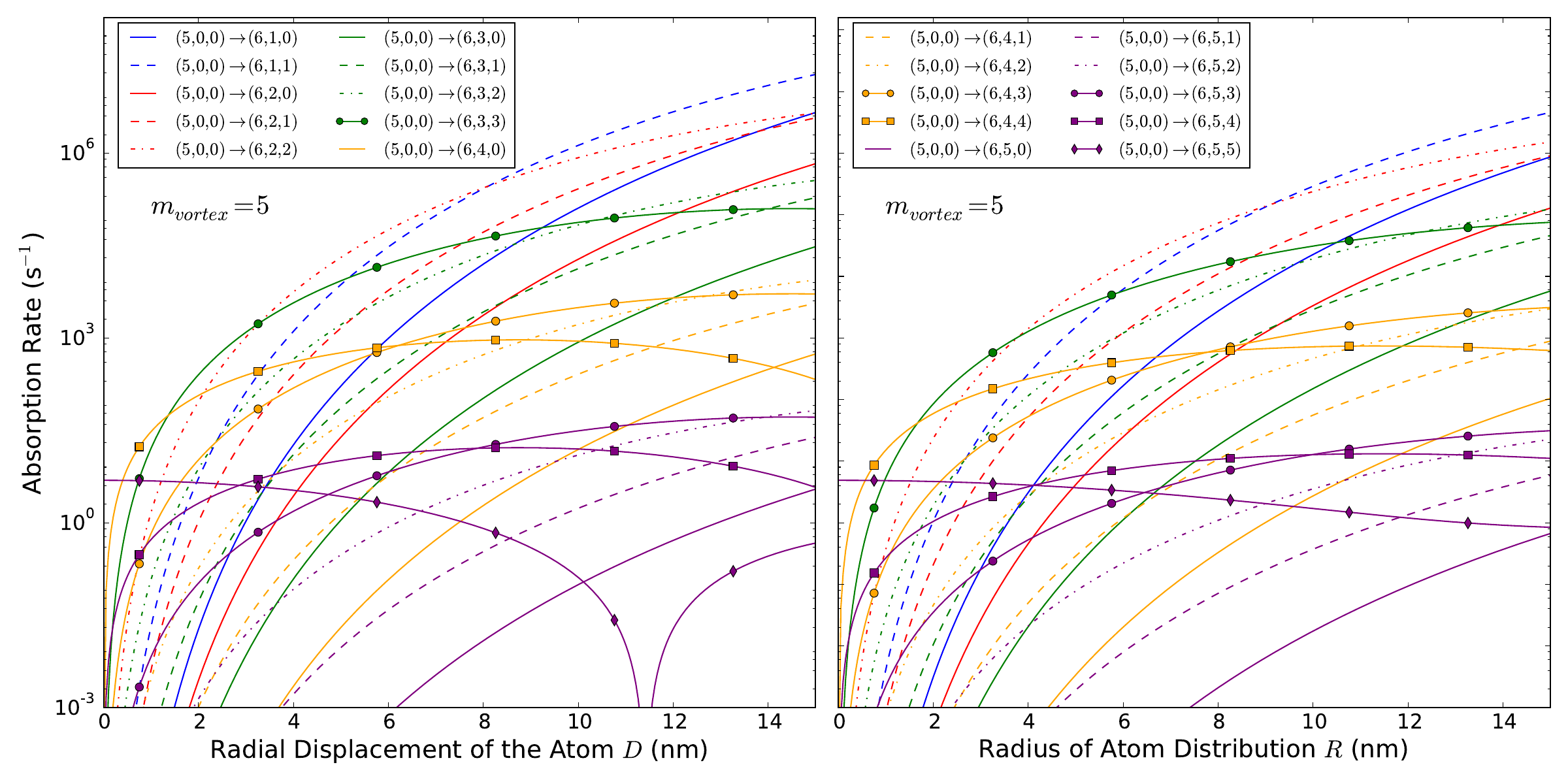}
    \caption{{\bf Single photon absorption rate of a vortex mode with OAM of $5\hbar$ between the state $(5,0,0)$ and all states in the family $(6,l,m)$ with $z_0 = 20\unit{nm}$ and $\eta = 250$.} This plot is based on \reffig{fig4_old} but includes all possible transitions to states in $(6,l,m)$. By comparing transitions to the same value of $m$ (same line style) we observe the same radial dependence up to an overall constant. This arises from the same Bessel enhancement term in \req{3} of the main text, but different baseline transition rates $\Gamma_{\Delta m}(i\to f,0)$. When comparing transitions to the same value of $l$ (same color), where the baseline transition for different $m$ is more comparable, we observe that the Bessel enhancement term induces a very different radial dependence. The competition between the Bessel enhancement term and the baseline transition enables the usage of the radius as a tuning parameter of the desired transition.}
    \label{fig:Fig3Full}
\end{figure}

Here, we provide the results of numerical calculations of absorption rates (normalized by the number of photons) for transitions between principal quantum number $5$ and $6$ in hydrogen. This provides more examples of the order of magnitudes of transition rates in the hydrogen atom as a function of multipolarity, z-projected angular momentum change, and displacement of the atom from the vortex center. In particular, in \reffig{fig4_old} and \reffig{fig:Fig3Full} we consider absorption of a vortex polariton mode with $m_{\mathrm{vortex}}=5$ using all other parameters equal to those of \rfig{3} of the main text.

In \reffig{fig4_old} we consider only dominant processes for different radial displacements of the absorber. In the left panel we consider an absorber at a displacement $D$ of the vortex, while on the right panel we consider a uniform probability distribution for the position of the absorber with radius $R$. We observe a qualitatively similar behavior as in \rfig{3}, where the placement of the absorber provides a tuning parameter for controlling which transition is dominant.

Having considered only the dominant transition processes in \reffig{fig4_old}, in \reffig{fig:Fig3Full} we consider all the different absorption processes possible when absorbing an $m_{vortex} = 5$ vortex mode. In particular we would like to focus on the relative strength of the different transitions. From \req{3} of the main text, we can understand the dependence of the absorption rate up to a factor determined by the matrix element of the aligned vortex mode, $\Gamma_{\Delta m}(i\to f, 0)$. Let us define $\Gamma_{\Delta m}(i\to f, 0)$ as the baseline transition rate. The form of \req{3} can be seen in these results, where transitions with equal $\Delta m$ have the same radial functional form, and are proportional to one another. This arises from different baseline transitions. Another trend we observe is that for equal $\Delta m$, the transition with the lowest $\Delta l$ has a greater rate, arising from a larger baseline transition rate.

\subsection{Effect of the Confinement factor on the Absorption rates of Polaritonic modes}

\begin{figure}
  \centering
  \includegraphics[width = 1\textwidth]{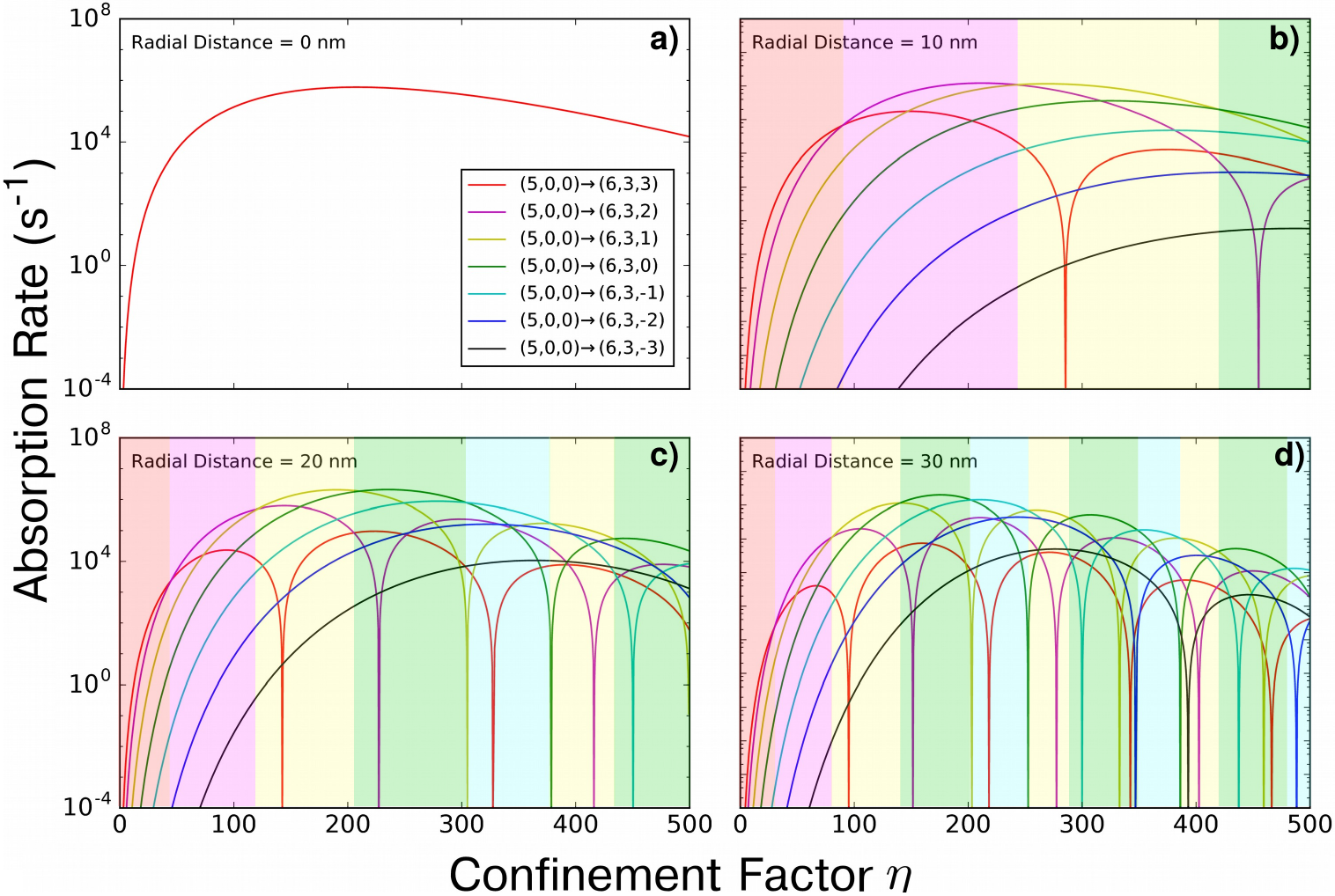}
  \caption{{\bf Impact of confinement factor on the absorption rate of a SP(h)P mode.} For different values of displacement distance (0,10,20,30 nm) the absorption rate of the family of transitions $(5,0,0) \to (6,3,\Delta m)$ for a hydrogen atom at $z_0 = 20$ nm is plotted as a function of the confinement factor $\eta$ of the SP(h)P mode. By tuning the confinement factor of a material, as in the case of graphene, it is possible to use a single experimental realization to access a variety of conventionally forbidden transitions, corresponding to the differently shaded areas, whose color matches the dominant transition. The confinement factor also determines the overall absorption of the vortex modes by matching the polaritonic and the electronic length scales, as discussed in the main text and \cite{Rivera2016}. The exponential decay of the vortex mode in the out of plane direction leads to an exponential decay of the absorption rate for large confinement leading to a maximum absorption rate for a system, as seen in the different panels.  The zeroes of the rates will be regulated by losses, which lead to an averaging over wavevectors.}
  \label{fig5_Sup}
\end{figure}

In \rfig{4} in the main text, we have analyzed what the dominant transition is for a particular value of confinement factor and displacement of the atomic system, given a specific vortex mode ($m_{vortex}=3$). In \reffig{fig5_Sup} we plot line cuts (along displacement) of \rfig{4} in order to more quantitatively illustrate the impact of the confinement factor on the absorption rates of the system. There are two particular features \reffig{fig5_Sup} makes clear. First, the existence of zeros in the absorption rates due to a change in the wavenumber scale in \req{3}. Second, the existence of competing mechanisms that can dominate the process: the exponential suppression of the field away from the dielectric surface, and the polynomial enhancement that arises from bridging the mismatch between the electromagnetic and electronic length scales. This latter feature is evident in the maximum in the absorption rate for a value of the confinement $\eta \approx 200$, for an atomic system at $z_0=20$ nm.

\begin{figure}[H]
  \centering
  \includegraphics[width = 0.8\textwidth]{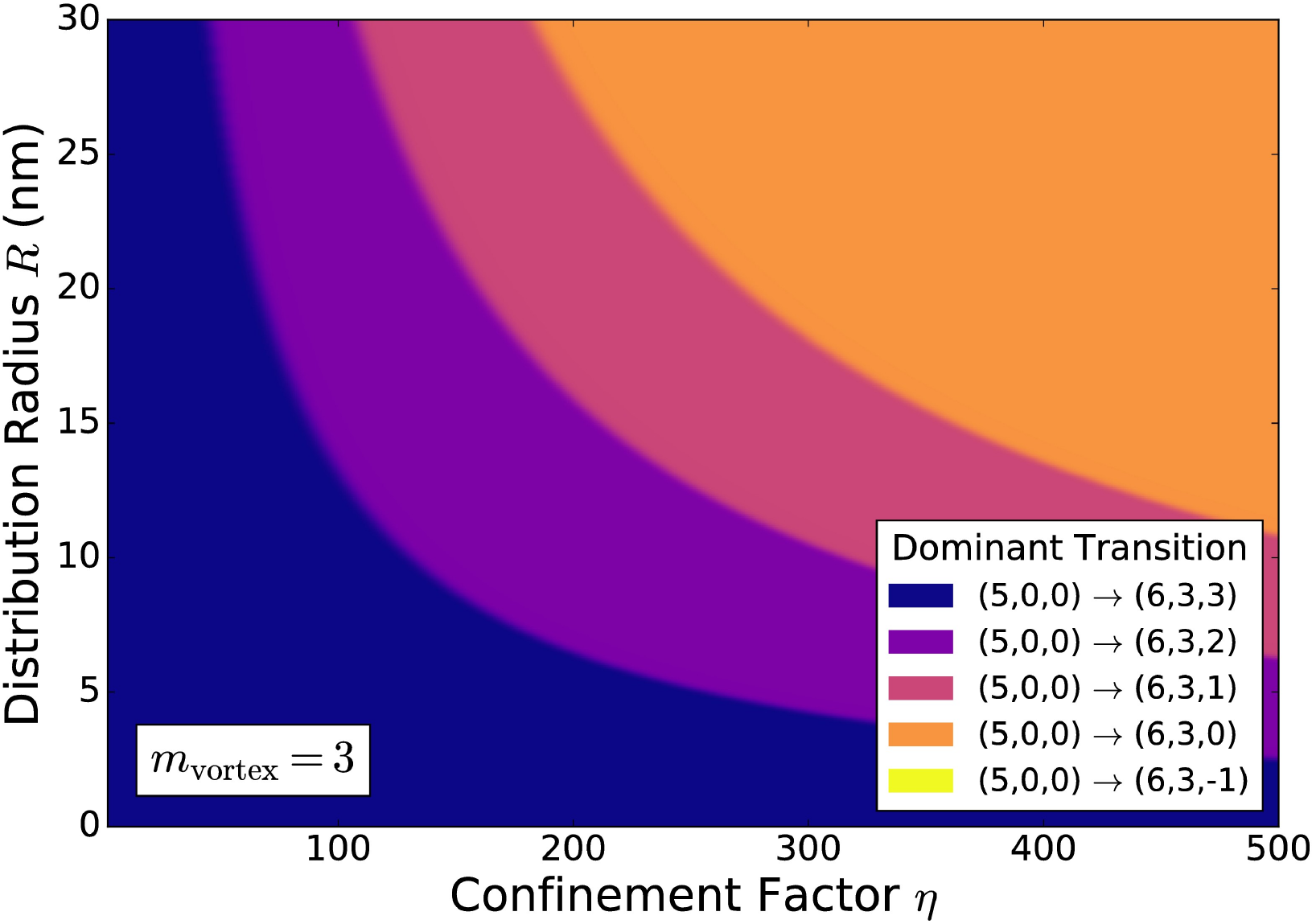}
  \caption{{\bf Landscape of dominant transitions for $m_{\text{vortex}} = 3$ for a distribution of atoms.} For distribution hydrogen atoms at $z_0=20$ nm, the dominant process in the family $(5,0,0)\to(6,3,\Delta m)$ is plotted as a function of the confinement factor $\eta$ and the radius $R$ of the uniform probability distribution of position. For a given vortex mode, the displacement and confinement can be used to control the dynamics of the electronic system even when there is uncertainty in the placement of the electronic system.}
  \label{fig4_Dist}
\end{figure}

In \reffig{fig4_Dist}, we extend the analysis done in \rfig{4} to the case of a probability distribution over the position of the absorbing atom. We compute the landscape of dominant transitions as a function of the confinement factor and the radius of the atom distribution. Similar to what we observed in \rfig{3}, the region of dominance of the different transitions are pushed to higher radius. Moreover, a transition becomes dominant for a larger set of parameters if its baseline transitions, $\Gamma_{\Delta m}(i\to f, 0)$ in \req{3}, is larger to begin with. This phenomena explains why $\Delta m=0$ transitions dominate for high displacement and confinement factor.

\begin{figure}
    \centering
  \includegraphics[width = 0.8\textwidth]{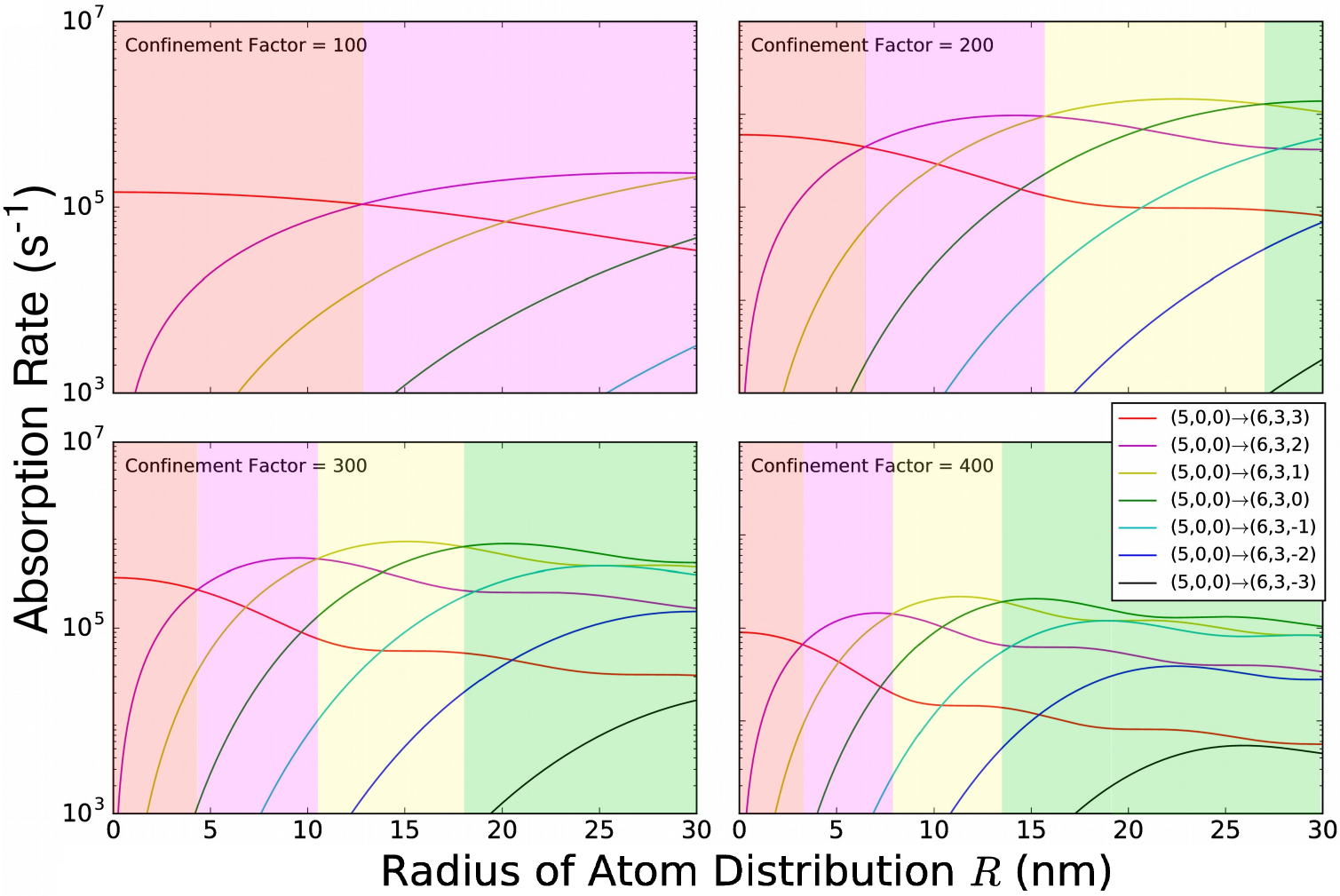}
  \caption{{\bf Impact of radius of atom distribution for different confinement factors on the absorption rate of a SP(h)P mode.} For a hydrogen atom at $z_0=20$ nm and for different values of confinement factor (100,200,300,400) the rate of absorption of the family of transitions $(5,0,0) \to (6,3,\Delta m)$ due to a $m_{\text{vortes}}=3$ SP(h)P mode is plotted as a function of the radius of the atom distribution. Although at $R=0$ zero we observe the angular momentum selection rules in \req{2}, for larger radii other transitions can become dominant, similar to \rfig{3}(e-h). As one increases the confinement factor, the length scale of the system decreases, leading to a decrease in the range where the different transitions dominate. Because of the averaging due to the distribution of atoms, there are no nodes in the absorption rates and for large radius of atom distribution the transition to $\Delta m =0$ always dominates.}
  \label{fig_R_cuts}
\end{figure}

\begin{figure}
  \centering
  \includegraphics[width = 0.8\textwidth]{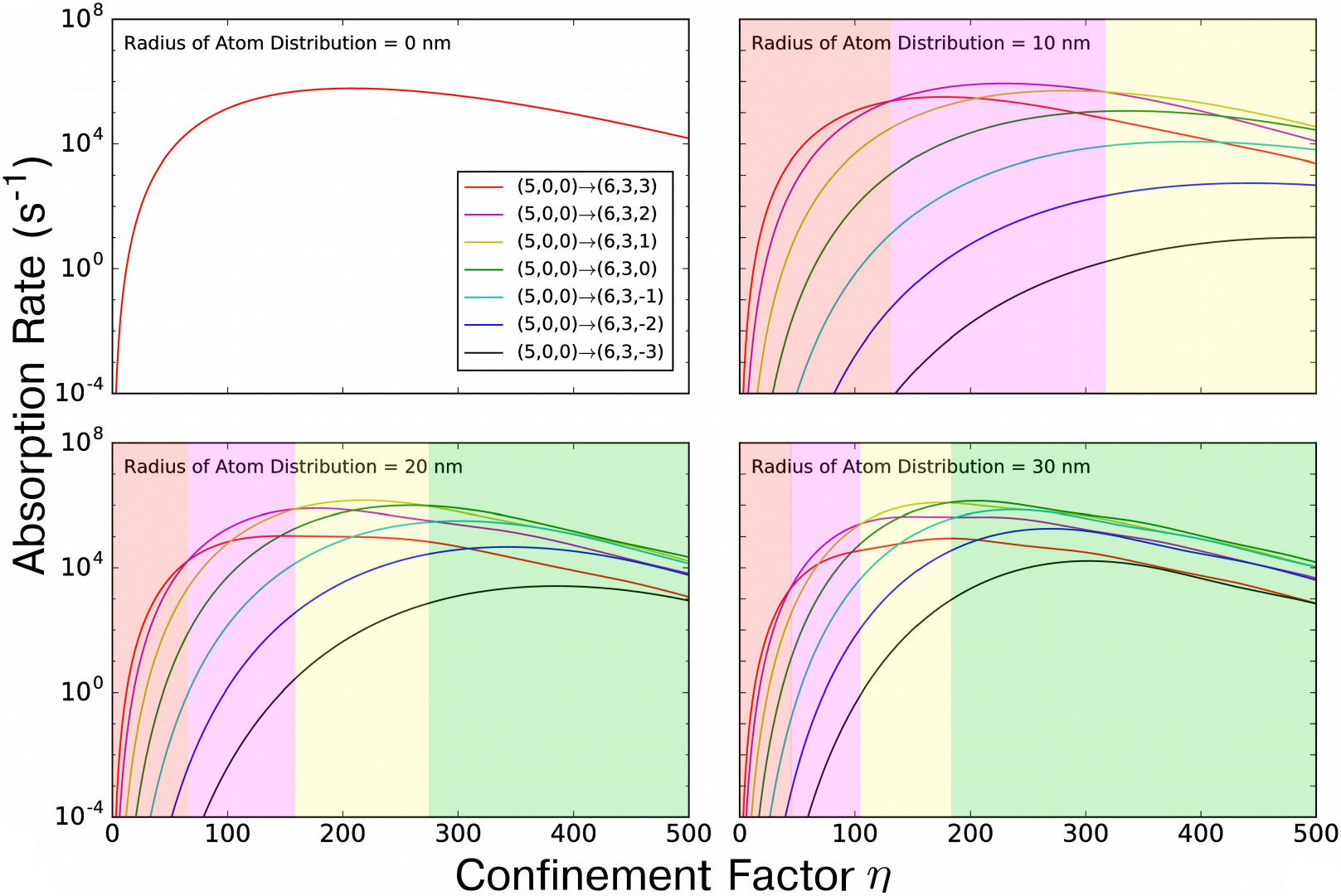}
  \caption{{\bf Impact of confinement factor for different radii of atom distribution on the absorption rate of a SP(h)P mode.} For a hydrogen atom at $z_0=20$ nm and for different values of radii of atom distribution (0, 10, 20, 30 nm) the rate of absorption of the family of transitions $(5,0,0) \to (6,3,\Delta m)$ due to a $m_{\text{vortes}}=3$ SP(h)P mode is plotted as a function of the radius of the atom distribution. When the radius of the atom distribution is zero, the selection rules in \req{2} is again satisfied. As the radius of the atom distribution increases the range of confinement where different transitions dominate decreases. This is a similar behavior to that in \reffig{fig_R_cuts} since radius and confinement factor play an equivalent role in the Bessel term in \req{3}. Again the role of the averaging removes any nodes in the absorption rate and leads to the transition $\Delta m=0$ dominating for large confinement factor, in contrast with the case of single particle location case summarized in \reffig{fig5_Sup}.
      }
  \label{fig_eta_cuts}
\end{figure}

To better understand the effect of the confinement when there is uncertainty in the placement of the absorber we plot the absorption rate fixing the confinement factor (\reffig{fig_R_cuts}) or the radius of the atom distribution (\reffig{fig_eta_cuts}).

  In \reffig{fig_R_cuts}, we look at a fixed confinement factor $\eta$ for $\eta = \{100,200,300,400\}$. In this case, the results are similar to those seen in \rfig{3}(e-h), where the radius of the distribution provides a good parameter to select which transition one wishes to make dominant. Looking at the different panels, it becomes clear that as one increases the confinement factor, the length scale of the polariton becomes smaller, leading to smaller ranges where the different transitions are dominant.

  In the different panels of \reffig{fig_eta_cuts}, we fix the radial displacement of the atom and vary the confinement factor. Again we observe that the overall transition rates are not monotonic with increasing confinement factor, in agreement with the discussion  of \reffig{fig5_Sup}. Unlike \reffig{fig5_Sup}, there are no nodes in \reffig{fig_eta_cuts} because of the averaging that occurs in the atom distribution. Also we note that for larger values of the confinement factor there is a smaller range of radii of atom distributions where each transition dominates. This is similar in \reffig{fig_R_cuts}, thus confirming the analogous role played by displacement and wavenumber in \req{3}.

\section{Appendix 2: Construction of Classical Polaritonic Vortex Modes in (An)isotropic Dielectrics}

In this section we describe classical surface polariton plane-wave modes for the case of a thin metallic film surrounded by dielectric. Through superposition of these planar modes we build vortex modes labeled by the winding number of the phase. We then prove their orthogonality and show how translated vortex modes can be resolved in a superposition of non-translated (centered) vortex modes. We then develop a simpler formalism for their construction, one that works when the modes are highly sub-wavelength (i.e., when the modes are nearly longitudinal (curl-free) and thus well-described by a scalar electric potential).

\subsection{General Construction}

Surface polariton modes generically correspond to excitations in solids which are coupled to electromagnetic modes highly confined to the interface of the material. Of particular interest are surface plasmon polaritons (SPP) and surface phonon polaritons (SPhP), where the solid excitation corresponds to a plasmon or phonon respectively. But other surface polaritons exist in nature, such as surface exciton polaritons in semiconductors like ZnO, and CuCl. The electromagnetic modes associated with these surface polaritons can be obtained by solving Maxwell's equations with the particular dielectric function (which may be spatially non-local) of the polaritonic material. For a thin film (quasi-2D) geometry with surrounding dielectrics (as in Figure 1 of the main text), the vector potential of the planar SP(h)P mode is given by \cite{jablan2009plasmonics}:
\begin{equation}\label{eq:Planar_SP(h)P}
  \bm{A}_{\bm{q}, \bm{\kappa}} = \frac{A_0}{\sqrt{q^2 + \kappa^2}} (\kappa\hat{q} + i q\hat{z}) e^{-\kappa z} e^{i(\bm{q}.\bm{\rho} - \omega t)} \quad \underset{\text{electrostatic limit}}{\Rightarrow} \quad {\bm A}_{\bm{q}} = \frac{A_0}{\sqrt{2}} (\hat{q} + i \hat{z}) e^{-q z} e^{i(\bm{q}.\bm{\rho} - \omega t)} \quad (z>0),
\end{equation}

\noindent where $\bm{q}$ is the in-plane momentum, $1/\kappa$ is the out of plane decay length, $z$ is the out of plane coordinate, $\bm{\rho}$ is the in-plane position, $\omega$ is the frequency of the mode, $A_0$ is the amplitude of the mode and hats represent unit vectors. Within the electrostatic limit, $q \gg \omega/c$, the inverse decay length of the mode equals the in-plane momentum, $\kappa \approx q$, enabling the approximation in Equation (S1) and \req{1} in the main text.

Having described the plane-wave surface polariton mode, we now build the vortex polariton modes as a superposition of surface plane-wave modes whose superposition coefficients are phases ($e^{im\alpha}$) proportional to the angle ($\alpha$) of the direction of the plane-wave mode with respect to some reference direction. In order for the mode to be well defined, as one goes around the vortex center, the phase of the mode must change by $2\pi m$ for $m\in\mathbb{Z}$. The value of $m$ corresponds to the winding number of the phase around the vortex center and determines its angular momentum $\hbar m$. When we discuss selection rules for electronic transitions involving the emission and absorption of vortex polaritons, we will be in a position to justify the interpretation of $\hbar m$ as angular momentum. The vortex mode is given by:

\begin{equation} \label{eq:VortexDef}
  \bm{A}_{q,m} = \frac{1}{2\pi}\int_0^{2\pi} d\alpha\; \frac{A_0}{\sqrt{2}} e^{i(\bm{q}.\bm{\rho} - \omega t)} e^{i\alpha m}
  \begin{cases} (\hat{q} + i \hat{z}) e^{-q z}  &  (z>0)\\
    \hat{q} & (z=0)\\
    (\hat{q} - i \hat{z}) e^{qz}  &  (z<0)
  \end{cases} ~,
\end{equation}

\noindent where  $\alpha$ is angle between $\bm{q}$ and some reference direction, set to be $\hat{x}$ without loss of generality. Let $\phi$ be the angle between $\bm{\rho}$ and $\hat{x}$. Using the Jacobi-Anger expansion \cite{NIST:DLMF} and the orthogonality of complex exponentials, we obtain, for $z>0$:
\begin{align}
 \bm{A}_{q,m}  =& \frac{A_0}{2\sqrt{2}\pi}e^{-qz}\int_{0}^{2\pi} d\alpha ~ (\cos(\alpha)\hat{x} + \sin(\alpha)\hat{y} + i \hat{z})  e^{iq\rho\cos(\alpha - \phi)} e^{i\alpha m}  \notag \\
 =& \frac{A_0}{4\sqrt{2}\pi}e^{-qz}\int d\alpha  ~ ((e^{i\alpha} + e^{-i\alpha})\hat{x} - i(e^{i\alpha} - e^{-i\alpha})\hat{y} + i 2\hat{z}) \sum_{n=-\infty}^{\infty} i^n J_n(q\rho) e^{in (\alpha - \phi)} e^{i\alpha m}  \notag \\
 =&  \frac{A_0}{2\sqrt{2}} e^{-qz}e^{im\phi} i^{m+1} \Big\{  (\hat{x} - i\hat{y})  J_{m+1}(q\rho) e^{i\phi} - (\hat{x} + i\hat{y}) J_{m-1}(q\rho) e^{-i\phi} +  \hat{z}2J_{m}(q\rho)\Big\} ~,  \notag
 \end{align}
 which implies more generally that
\begin{equation}
\bm{A}_{q,m}= e^{-q|z|}  \frac{A_0}{2\sqrt{2}} e^{im\phi} i^{m+1} \Bigg(  \hat{\rho} [J_{m+1}(q\rho) - J_{m-1}(q\rho)] -i \hat{\phi} \frac{2m J_{m}(q\rho)}{q\rho} +
 \left. \begin{cases}
   \hat{z}2J_{m}(q\rho) & (z>0)\\
   0 & (z=0)\\
   - \hat{z}2J_{m}(q\rho) & (z<0)
 \end{cases}\right). \label{eq:CylindricalForm}
\end{equation}

\subsection{Energy and Orthogonality of Vortex Modes}

Consider the following inner product between two modes (which is proportional to the time-averaged energy density of the mode):
\begin{equation}\label{eq:innerProduct}
\langle \bm{A}_{q,n}, \bm{A}_{k,m}\rangle = \int d\bm{r} \frac{1}{2\omega} \frac{d(\epsilon_r\omega^2)}{d\omega}   \bm{A}_{q,n}^*(\bm{r}) \bm{A}_{k,m}(\bm{r}).
\end{equation}
For the purposes of this work we consider the substrate and superstrate to be negligibly dispersive in the frequency range of interest ($\frac{\partial\epsilon}{\partial\omega} \ll \frac{\epsilon}{\omega}$). For the case of 2D plasmons, we consider the Drude model for the dielectric response of a metal, given by:
\begin{equation}
  \epsilon_{r,metal}(\omega) =  1 - \frac{\omega_p^2}{\omega^2},
\end{equation}
where $\omega_p$ is the plasmon frequency of the metal. For the integration of the thin metallic film volume, we take a finite width slab and take the limit as the thickness $d$ goes to zero. The dielectric function can then be written as:
\begin{equation}
  \epsilon_{r,film} = \left(1 - \frac{\omega_p^2}{\omega^2}\right)\theta(d/2 - |z|)  \quad\underset{d\to0}{\Rightarrow}  \quad\epsilon_{r,film} = \left(1 - \frac{\omega_p^2}{\omega^2}\right)d\delta(z) \nonumber,
\end{equation}
where $\theta(x)$ is a step function. The derivative term will cancel the $\omega_p^2/\omega^2$ terms, while the first term is zero in the limit $d\to 0$. The inner product then simplifies to (for the Drude plasmons with high confinement):
\begin{equation}
  \langle \bm{A}_{q,n}, \bm{A}_{k,m}\rangle = 2\bar{\epsilon_r} \int\limits_{z>0} d\bm{r}  \bm{A}_{q,n}^*(\bm{r}) \bm{A}_{k,m}(\bm{r}), \nonumber
\end{equation}
where $\bar{\epsilon}_r$ is the average dielectric surrounding the metallic film. By definition this integral is:
\begin{align*}
  \langle \bm{A}_{q,n}, \bm{A}_{k,m}\rangle = \bar{\epsilon}_r \frac{A_0^2}{2}  \int_0^\infty dz & \int_0^\infty d\rho\;\rho \int_0^{2\pi} d\phi\; e^{-z(q+k)} e^{i(m-n)\phi} \times\\
  \times &\left[ J_{m+1}(k\rho) J_{n+1}(q\rho) + J_{m-1}(k\rho) J_{n-1}(k\rho) + 2J_m(k\rho)J_n(k\rho)\right] 
\end{align*}
Since the $\phi$ integration imposes $m=n$, the Bessel functions differ only by their argument $k\rho$ or $q\rho$. Using the orthogonality of Bessel functions, i.e., $\int d\rho~\rho J_m(k\rho)J_m(q\rho) = \frac{1}{k}\delta(k-q)$, the inner product vanishes for $k\neq q$, so the modes are orthogonal. For $k=q$, we can evaluate the inner product (proportional to "$\delta(0)$") by setting the upper limit of the radial integral to a normalization length (also often called a quantization length), $L$, much larger than any length scale associated with the plasmonic vortex. Any result we will derive will be independent of $L$. Setting the upper limit of the integral to $L$, the inner product for $k=q$ is simply $\frac{L}{\pi q}$.

\subsubsection{Angular Momentum Representation of Displaced Vortices}

To define a vortex mode displaced from our coordinate center by $\bm{D}$ we consider the translation of \refeqP{eq:VortexDef} by $\bm{D}$ (setting $\boldsymbol{\rho}$ to $\boldsymbol{\rho}-\boldsymbol{D}$):
\begin{equation} \label{eq:DispVortexDef}
  \bm{A}_{q,m}^{\bm{D}} = \frac{1}{2\pi}\int d\alpha  \frac{A_0}{\sqrt{2}} (\hat{q} + i \hat{z}) e^{-qz} e^{i(\bm{q}.\bm{\rho} - \omega t)} e^{-i\bm{q}\cdot \bm{D}} e^{i\alpha m}  \quad (z>0).
\end{equation}
Applying the Jacobi-Anger relation to the term $e^{-i\bm{q}\cdot \bm{D}}$ one obtains:
\begin{equation}
  e^{-i\bm{q}\cdot \bm{D}} = e^{-iqD\cos( \alpha - \phi_0)} = \sum_{n=-\infty}^\infty (-i)^n J_{n}(qD) e^{in (\alpha - \phi_0)},
\end{equation}
\noindent where $\phi_0$ is the angle between $\bm{D}$ and the chosen reference direction ($\hat{x}$ here). The displaced vortex mode can then be easily written as:
\begin{align}
  \bm{A}_{q,m}^{\bm{D}} &= \sum_{n=-\infty}^\infty (-i)^n J_{n}(qD) e^{-in\phi_0} \frac{1}{2\pi}\int d\alpha  \frac{A_0}{\sqrt{2}} (\hat{q} + i \hat{z}) e^{-qz} e^{i(\bm{q}.\bm{\rho} - \omega t)} e^{i\alpha (m + n)}  \quad,\quad (z>0) \notag \\
  &= \sum_{n=-\infty}^\infty (-i)^{n-m} J_{n-m}(qD) e^{-i(n-m)\phi_0} \bm{A}_{q,n}.
\end{align}
The displaced vortex can thus be expressed as a superposition of vortex modes centered around the coordinate system. This fact enables us to reduce the displaced vortex mode case to a superposition over centered vortex modes, whose computation and understanding is much easier due to the symmetry of the mode.

\subsection{Quasi-Electrostatic Limit: Scalar Potential Formalism}

We now discuss an alternative and simpler method for computing the vortices which applies for fields which are approximately described by the gradient of a scalar potential (i.e., approximately longitudinal) \footnote{Note that this does \textit{not} mean that we have already switched from the $\phi = 0$ gauge to the Coulomb gauge, where the vector potential is approximately zero. We merely mean that the mathematical form of the electric field or vector potential modes in the $\phi = 0$ gauge is the gradient of a scalar function (i.e., the field is approximately curl-free). This is the real-space definition of field longitudinality. In general, a purely longitudinal field can indeed be represented by having zero scalar potential and a non-zero but longitudinal vector potential.}. As can be seen from the plane-wave polariton modes in the electrostatic limit, they can indeed be written as the gradient of a scalar potential. By the linearity of derivatives, the vortex modes constructed from these plane-wave modes can also be expressed as gradients of a scalar potential. We now focus on constructing these vortices for \textit{any} uniaxial or isotropic medium. We include the possibility of uniaxial anisotropy due to the recent interest in the phonon-polaritons of hBN. To find the modes, we solve the Laplace equation subject to appropriate boundary conditions. The Laplace equations in each region read (using repeated index notation):
\begin{equation}
\partial_i \epsilon_{ij} \partial_j\phi =  0 \nonumber.
\end{equation}
The translational invariance in-plane admits scalar potential solutions of the form $Z(z)e^{i\mathbf{q}\cdot\boldsymbol{\rho}}$. Taking the optical axis of a uniaxial crystal to be perpendicular to the slab illustrated in Figure 1 of the main text, the dielectric function of such a uniaxial crystal can be written as $\epsilon(\omega) = \text{diag}(\epsilon_{\perp}(\omega),\epsilon_{\perp}(\omega),\epsilon_{||}(\omega))$. In that case, the Laplace equation becomes:
\begin{equation}
-\frac{\epsilon_{\perp}}{\epsilon_{||}}q^2Z + \frac{d^2Z}{dz^2} = 0 \implies \frac{d^2Z}{dz^2}=r^2q^2Z \nonumber,
\end{equation}
where the anisotropy ratio, $r$ is defined as $\sqrt{\frac{\epsilon_{\perp}}{\epsilon_{||}}}$.  The scalar potential vortex mode is thus defined as:
\begin{equation}
\phi_{q,m} \equiv \frac{1}{\sqrt{2}q}N_{q,m}Z_q(z)\int\limits_0^{2\pi} \frac{d\alpha}{2\pi} \  e^{i\alpha m+iq\rho\cos(\alpha-\phi)},
\end{equation}
where $N_{q,m}$ is a normalization constant to be specified during quantization. Due to the in-plane rotational symmetry, $Z_q(z)$ is independent of the angle that the wavevector makes to the reference direction. Via the Jacobi-Anger expansion, this previous expression becomes
\begin{equation}
\phi_{q,m} = \frac{1}{\sqrt{2}q}i^mN_{q,m}Z_q(z)J_m(q\rho)e^{im\phi}.
\end{equation}
The electric field (in either $\phi =0$ gauge or Coulomb gauge) and vector potential modes (in the Coulomb gauge) must be proportional to the gradient of this scalar potential mode in the electrostatic limit. The electric field vortices are therefore:
\begin{equation}
\mathbf{E}_{q,m} =-\nabla\phi_{q,m} =  -\frac{1}{\sqrt{2}q}i^mN_{q,m}e^{im\phi}Z_q\left[\frac{\partial J_m(q\rho)}{\partial\rho}\hat{\rho} + \frac{im}{\rho}J_m(q\rho)\hat{\phi} + \frac{1}{Z_q}\frac{dZ_q}{dz}J_m(q\rho)\hat{z}   \right] .\nonumber
\end{equation}
The vector potentials (in the $\phi=0$ gauge) are just $\frac{1}{i\omega}\mathbf{E}$. Using derivative relations for Bessel functions, the above equation can be expressed as:
\begin{equation}
\mathbf{E}_{q,m} =  \frac{1}{2\sqrt{2}}i^mN_{q,m}e^{im\phi}Z_q\left[[J_{m+1}(q\rho)-J_{m-1}(q\rho)]\hat{\rho} - \frac{2im}{q\rho}J_m(q\rho)\hat{\phi} - 2\frac{1}{qZ_q}\frac{dZ_q}{dz}J_m(q\rho)\hat{z}   \right].
\end{equation}
We pause to note that taking $Z_q(z) \sim e^{-q|z|}$ for graphene yields a field proportional to \req{5} and \req{S3}, confirming that the vortices can also be derived from a scalar potential vortex $-$ no surprise given the high confinement of the modes. Therefore, this approach very straightforwardly allows for the calculation of quasielectrostatic vortices in any 2D-translationally-invariant photonic system. One simply needs the appropriate $Z_q(z)$. To deal with isotropic systems, one need simply set $\epsilon_{\perp}$ to $\epsilon_{||}$ from the beginning. 

We conclude this section by noting that the electrostatic limit leads to remarkably simple expressions, despite the anisotropy of the system. It is important that we presented both a formalism which did not crucially rely on the electrostatic potentials (computing the vector potential of the mode, accounting for retardation), and a formalism which takes the electrostatic limit as its primary assumption. Obviously, the latter formalism is much simpler computationally. That said, the first formalism can be quite useful for plasmons which are not so strongly confined, as in the case of plasmons on top of a thick film of noble metal like silver or gold. The latter formalism is also useful for OAM carrying whispering gallery modes in micro-ring resonators. For problems in which relatively large emitters like quantum dots are interfaced with these electromagnetic modes, an approach ignoring retardation effects would be questionable, and for this reason, we have provided a formalism by which to understand the interaction between matter and vortices, even outside of the electrostatic limit.

\section{Appendix 3: Construction of Quantum Fields for Polaritonic Vortices in the Angular Momentum Basis}

\subsection{Field Operators}

Here, we derive the quantized electromagnetic field operators needed to describe interactions between electrons and polaritons. Since the main interest of our paper is vortex-enabled light-matter interactions, we will write the field operators in the angular momentum basis. Moreover, since the most interesting situations arise when the electromagnetic fields are approximately longitudinal, we will focus on those in this section. The extension to fields which are not quasi-electrostatic is straightforward.  In the first part, we will derive the form of the vector potential operator in the $\phi=0$ gauge, whose use can be extended to non-quasistatic vortices. Then we will present (just like in the last section) an alternative and simpler method of calculation via scalar potentials which can be used in the Coulomb gauge.
\subsubsection{Vector Potential in the $\phi=0$ Gauge}
We will now find the normalization of these modes that makes them suitable for QED calculations. The prescription for writing down field operators such as the vector potential operator is to write it as \cite{Glauber1991}:
\begin{equation}
\mathbf{A} = \sum_{q,m} \sqrt{\frac{\hbar}{2\epsilon_0\omega_{q,m}}} (a_{q,m}\boldsymbol{F}_{q,m}+h.c),
\end{equation}
where the $\boldsymbol{F}_{q,m}$  are proportional to \textbf{E}$_{q,m}$ but normalized such that $\int \frac{1}{2\omega}\boldsymbol{F}^*_{q,m}\cdot\frac{d(\epsilon\omega^2)}{d\omega}\cdot\boldsymbol{F}_{q,m} =1$. The double dot product is present because of the regions of anisotropic dielectric. The energy integral takes the form:
\begin{equation}
\frac{1}{2\omega}\int dz\  d\phi\  d\rho~ \rho\left[ \frac{d(\epsilon_{\perp}\omega^2)}{d\omega}\left(|F^x_{q,m}|^2+|F^y_{q,m}|^2\right) + \frac{d(\epsilon_{||}\omega^2)}{d\omega}|F^z_{q,m}|^2\right]. \nonumber
\end{equation}
Taking $\boldsymbol{F}_{q,m}$ to be given by \req{S15} and performing the azimuthal integral first yields a factor of $2\pi$. Performing the radial integral, making use of the orthogonality of Bessel functions yields (via the same arguments as those used when we computed the energy of graphene vortices in Sec. III.B):
\begin{equation}
\frac{2|N_{q,m}|^2L\bar{\epsilon}_r\xi_{q}}{q^2}, \nonumber
\end{equation}
where $\xi_{q}$ is dimensionless and given by:
\begin{equation}
\xi_{q} = \frac{q}{4\omega\bar{\epsilon}_r}\int dz~\left[\frac{d(\epsilon_{\perp}\omega^2)}{d\omega}|Z|^2 + \frac{d(\epsilon_{||}\omega^2)}{d\omega} \frac{1}{q^2}\Big|\frac{dZ}{dz}\Big|^2 \right].
\end{equation}
We therefore see that the vector potential is given by:
\begin{equation}
\mathbf{A} = \sum\limits_{q,m}\sqrt{\frac{\hbar q^2}{4\epsilon_0\bar{\epsilon}_rL\omega_{q,m}\xi_{q}}}(a_{q,m}\boldsymbol{F}_{q,m}+h.c),
\end{equation}
where
\begin{equation}
\boldsymbol{F}_{q,m} = \frac{1}{2\sqrt{2}}i^{m}e^{im\phi}Z(z)\left([J_{m+1}(q\rho)-J_{m-1}(q\rho)]\hat{\rho} - \frac{2im}{q\rho}J_m(q\rho)\hat{\phi} - 2\frac{1}{qZ_q}\frac{dZ_q}{dz}J_m(q\rho)\hat{z}   \right).
\end{equation}
In the region $(z>d/2)$, we have that $Z_q(z)$ is just $e^{-qz}$. Therefore, the expression for $\boldsymbol{F}_{q,m}$ exactly matches \req{5}. We have defined $\xi_{q}$ in order to make the field operators for other polaritons strongly resemble those field operators for graphene. In fact, for graphene (or any 2D plasmonic material with Drude dispersion), $\xi_{q} = 1$. The electric field operator is derived from the vector potential in a gauge where the scalar potential is zero via $\mathbf{E} = i\omega\mathbf{A}(\omega)$ (this means that the h.c. term  picks up a relative minus sign since it is counter-rotating relative to the first term in the Heisenberg representation of the field operator).

\subsubsection{Scalar Potential in the Coulomb Gauge: $\nabla\cdot\mathbf{A}=0$}
Changing to Coulomb gauge and using the electrostatic limit, the vector potential vanishes everywhere. Now, all the dynamics are generated by a scalar potential. Using the potentials in \req{S10} with the derived normalization, the scalar potential operator can be written:
\begin{equation}
\phi = \sum_{q,m} \sqrt{\frac{\hbar\omega}{4\epsilon_0\bar{\epsilon}_r\xi_{q} L}}(a_{q,m}U_{q,m}+h.c.),
\end{equation}
where
\begin{equation}
U_{q,m} = \frac{1}{\sqrt{2}}i^mZ_q(z)J_m(q\rho)e^{im\phi}.
\end{equation}
\subsubsection{$\xi_{q}$ for Different Materials}
Our findings apply to all uniaxial media whose surface polaritons are well described by longitudinal electric fields. The difference between each material is contained in the $\xi_{q}$ factor and the ratio $c/v_g$. It applies to plasmons in graphene, to 2D plasmons in metallic monolayers, to phonon-polaritons in hBN or SiC \cite{dai2014tunable,caldwell2013low},  to the acoustic plasmons in gated graphene \cite{alonso2016ultra}, to surface exciton polaritons like in MoS$_2$ \cite{karanikolas2016near} and in newly discovered phase-change materials \cite{li2016reversible}, and could, with straightforward extension of the formalism, apply to anisotropic plasmons in black phosphorus \cite{low2014plasmons}. $\xi_{q}$ was defined for graphene such that $\xi_{q}(\text{graphene})=1$. The reason for this is that the rates of various electronic transitions in the vicinity of graphene take very simple analytic forms \cite{Rivera2016} (simpler than all of the other materials we mentioned in the text). That makes it easy to understand the order of magnitude of transitions in other materials by making graphene the reference material. 

Of course, all materials are different, and material-specific and emitter-specific challenges arise when interfacing particular emitters with particular materials. Because of this it is important to formulate our theory in a way that makes explicit that similar results hold for a wide range of material systems. It makes explicit the fact that from the perspective of photonics and transition rates,  all of these above-mentioned materials are similar. Thus, we hope that the unity of the treatment of different polaritonic materials will serve to provide a platter of options for realizing the predictions of the theory. 

In the rest of this section, we discuss explicitly (i.e., with numbers) the case of hyperbolic polaritons in a slab of hexagonal boron nitride of thickness $d$.  The electric field plane-wave modes can be shown to be:
\begin{equation}
  \bm{E}_{q,m} = \frac{E_0}{i\omega\epsilon_0}\epsilon_r^{-1}(z)\nabla\times \hat{q}_{\perp}
  \begin{cases}  e^{-qz}  &  (z>0)\\
    \frac{e^{ikd}\epsilon_{\perp}q(\epsilon_sk-i\epsilon_{\perp}q)}{2k(-\epsilon_{\perp}q\cos(kd)+\epsilon_{s}k\sin(kd))}e^{ikz}+\frac{\epsilon_{\perp}q(-i\epsilon_sk+\epsilon_{\perp}q)}{k(\epsilon_sk+i\epsilon_{\perp}q+e^{2ikd}(-\epsilon_{s}k+i\epsilon_{\perp}q))}e^{-ikz} & (-d<z<0)\\
    \frac{e^{qd}q\epsilon_{\perp}\epsilon_s}{-\epsilon_{\perp}q\cos(kd)+\epsilon_{s}k\sin(kd)} e^{qz}, &  (z<-d)
  \end{cases} 
\end{equation}
where $\epsilon_r$ is the relative permittivity of the hBN which is given by $\text{diag}(\epsilon_{\perp},\epsilon_{\perp},\epsilon_{||})$, $\epsilon_s$ is the relative permittivity of the substrate, $q$ is the polariton wavenumber, $k \equiv q\sqrt{r}$, $r = \Big|\frac{\epsilon_{\perp}}{\epsilon_{||}}\Big|$, $\hat{q}_{\perp}$ is a unit vector along the in-plane direction perpendicular to the wavevector, and $E_0$ is a normalization constant. The dispersion satisfies the following implicit equation:
\begin{equation}
\tan(kd) = \frac{1+\epsilon_s}{\frac{\epsilon_s\sqrt{r}}{\epsilon_{\perp}}-\frac{\epsilon_{\perp}}{\sqrt{r}}}.
\end{equation}
\begin{figure}
  \centering
    \includegraphics[width=0.7\textwidth]{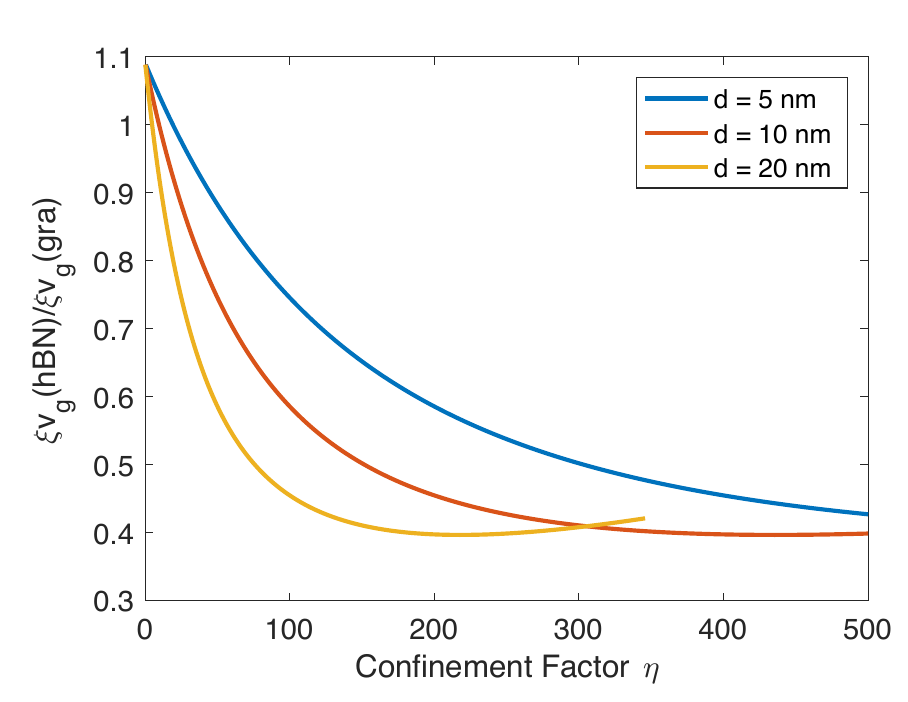}
    \caption{{\bf Analysis of the dimensionless quantity $\xi v_g$ that captures the choice and geometry of the material.} Comparison of the product $\xi_q v_g$ between hBN and graphene for hBN slabs of thicknesses 5, 10, and 20 nm. The comparison is done as a function of confinement in order to achieve a direct comparison between the strengths of light-matter interactions at fixed confinement factor. The plot is terminated at a maximum confinement of 500 (a polariton wavelength of about 10 nm), although the theory predicts potentially even higher confinements for the 5 and 10 nm thick hBN near the very edge of the Reststrahlen band. The yellow curve terminates at a confinement of 300 because the maximum frequency for which the confinement is computed is 1600 \text{cm}$^{-1}$.}
    \label{fig:S1}
\end{figure}

\noindent These two pieces of information suffice to compute the confinement, group velocity, and $\xi_q$ factors for hBN. In \rfig{S8}, we show how the factor $\xi_q v_g$ compares between hBN and graphene plasmons for hBN films of different thicknesses. We perform the computation in the upper Reststrahlen band. The only difference in emission and absorption rates of single vortex polaritons comes from these factors and yet the two factors are within the same order of magnitude, making it clear and explicit that hBN vortices can be used to tailor selection rules in atomic systems.

\section{Appendix 4: Angular Momentum and Light-Matter Interactions}

\subsection{First-Order Processes}
The transition rate for the absorption is calculated using Fermi's Golden Rule (for radiation with a frequency spectrum much broader than the linewidth of the excited state):
\begin{align}
  \Gamma &= \frac{2\pi}{\hbar} \left| \langle \phi_f, n-1 | H_{int} | \phi_i, n\rangle \right|^2 \rho_m(\hbar \omega)
\end{align}

\noindent where $\phi_i$ ($\phi_f$) is the electronic initial (final) state, $n$ corresponds to the number of photons in our mode of interest, $\rho_m(\hbar \omega)$ corresponds to the density of vortex modes at each angular momentum $\hbar m$ \footnote{The calculation scheme follows the treatments described in \cite{guery2011advances,craig1984molecular} In this scheme, the initial state is taken such that the initial state of the electromagnetic field represents a collection of photons in a continuum of modes \{q\} with respective photon number \{n(q)\}. The continuum of final states arises from all of the ways that one photon can be removed from this distribution. The delta function in Fermi's Golden Rule picks "the number of photons" at the frequency of interest. For a more complicated initial state representing a superposition of photon number distributions, the mean number of photons at the frequency of interest shows up instead. We put this in quotes because it is impossible to have a finite number of photons in each mode if the modes form a continuum. Such a state would have clearly infinite energy. However it is possible to define the photon number in a way that is in agreement with the treatment in which a semi-classical treatment of the field is used. We do so in what follows.}. $H_{int}$ is the interaction Hamiltonian given by $\frac{e}{2m}(\mathbf{p}\cdot\mathbf{A}+\mathbf{A}\cdot\mathbf{p}) + \frac{e^2}{2m}\mathbf{A}^2$ in the $\phi=0$ gauge and by $e\phi$ in the Coulomb gauge. To compute $\rho_m(\hbar \omega)$ consider a perfectly conducting cylinder of extremely large radius (quantization length) $L$ surrounding the vortex mode (for example, take $L = 1$ m). At $L$, the non-radial components of the field must vanish. Using the large distance asymptotic form of the Bessel functions, the $q$ spacing of allowed states is given by $\pi/L$. The density of states is then given by:
\begin{equation}
  \rho_m(\hbar\omega_0) = \frac{L}{\pi} \int dq \ \delta(\hbar \omega_0 - \hbar\omega) = \frac{L}{\pi\hbar} \frac{1}{|{\bm{v}_q}|} ,
\end{equation}
where $\bm{v}_q$ is the group velocity $\frac{d\omega}{dk}$ of the mode at energy $\omega_0$. Note that by definition, the $q$ in the vortex mode is $|\mathbf{q}|$ and is therefore positive. In the case of SPP, the dispersion relation is approximately given by \cite{jablan2009plasmonics} (where $\beta$ is a proportionality constant which will appear nowhere in the final answers below) :
$$
q = \beta\omega^2 \quad  \Rightarrow\quad  \frac{dq}{d\omega} = 2\beta \omega \quad \Rightarrow \quad v_q = \frac{1}{2\beta\omega} = \frac{c}{2\eta},
$$
where $\eta$ is the confinement factor $\frac{qc}{\omega}$, which is the ratio of photon and polariton wavelengths at the same frequency. The absorption rate of the system is given by (taking the vector potential interaction Hamiltonian):
\begin{align}
  \Gamma &=  \frac{2\pi}{\hbar^2} \frac{L}{\pi} \frac{2\eta}{c} \frac{e^2}{m^2} \frac{n\hbar q^2}{4\bar{\epsilon}_r\epsilon_0 L\omega_{q,m}\xi_q}   \left|\langle \phi_f, n-1 | \bm{F}_{q,m} \cdot \hat{p} | \phi_i, n\rangle \right|^2 \notag\\
  &=  \frac{4\pi e^2}{4\pi\epsilon_0\hbar c}   \frac{n\eta^3\omega}{\bar{\epsilon}_r\xi_q}   \frac{\left|\langle \phi_f, n-1 | \bm{F}_{q,m} \cdot \hat{p} | \phi_i, n\rangle \right|^2}{ m^2c^2} \notag\\
  &=   \frac{4\pi n\alpha\eta^3\omega}{\bar{\epsilon}_r\xi_q}   \frac{\left|\langle \phi_f | \bm{F}_{q,m} \cdot \hat{p} | \phi_i\rangle \right|^2}{ m^2c^2}.
\end{align}

The photon number $n$ is derived by applying the correspondence principle to a hypothetically monochromatic radiation field (relative to its central frequency, but still much broader than the atomic linewidth due to spontaneous emission or collisions). This follows the treatment in \cite{craig1984molecular}. Consider a classical field with energy density per unit frequency $\frac{dU}{d\omega}$. This should just be equal to the mean photon number at frequency $\omega$ ($\langle n(\omega)\rangle$) times $\hbar\omega$ times the density of states $\rho(\omega)$. In that case, the mean photon number is simply $\frac{1}{\rho\hbar\omega}\frac{dU}{d\omega},$ where $U$ is the energy of an approximately single mode with the field amplitude of the classical field in the experiment of interest\footnote{It is also worth noting that when the field contains many photons and can be treated semiclassically, the absorption rate can be shown to be the same as what we derive here given that we define photon number in the way shown above. The answer one would get by applying first-order time-dependent perturbation theory is $\Gamma = \frac{2\pi e^2}{\epsilon_0m^2\omega_0^2\hbar^2} \int d\mathbf{r}d\mathbf{r'}~ \left(\frac{d\langle u_{ij}\rangle}{d\omega}\Big|_{\omega_0}(\mathbf{r},\mathbf{r'})\right)\psi_f^*(\mathbf{r}) \psi_f(\mathbf{r'})p_ip^{*'}_j\psi_i(\mathbf{r}) \psi_i^*(\mathbf{r'})$, where $\left(\frac{d\langle u_{ij}\rangle}{d\omega}\Big|_{\omega_0}(\mathbf{r},\mathbf{r'})\right)$ is the time-averaged cross-spectral density evaluated at two space points. The time-averaged cross-spectral density is simply the time-average of $\epsilon_0E_i(\mathbf{r},\omega)E_j^*(\mathbf{r'},\omega)$, where $E_i(\mathbf{r},\omega)$ is the Fourier transform of the $i-$th component of the semiclassical electric field. This can be seen as a generalization of Einstein's theory of absorption in which the absorption rate is proportional to the power spectral density at the location of the atom. The power spectral density is the trace of the time-averaged cross-spectral density evaluated at the same space points. For non-dipole transitions, the dependence on energy density is more complicated as a result of the spatial extent of the electronic wavefunctions.}.

We also point out here what happens if for some reason, the radiation is narrower than the linewidth of the excited atomic state (either due to radiation or other broadening mechanisms). We'll consider only the case of radiative broadening. In that case, the fact that the excited state is actually a continuum of width $\Gamma_d$ is relevant (where $\Gamma_d$ is the decay rate of the excited atomic state). In such a case, Fermi's Golden Rule would give that the decay rate is (on resonance) roughly equal to $\frac{2\pi}{\hbar^2\Gamma_d}|\frac{e}{m}\mathbf{A}\cdot\mathbf{p}|_{eg}^2$, where $\mathbf{A}$ is the vector potential amplitude of the approximately monochromatic (and semiclassical) radiation \cite{guery2011advances}. Nevertheless, the same matrix elements appear as in the broadband case, meaning that angular momentum conservation still appears in the transition rates. Therefore, regardless of whether the vortex (of fixed angular momentum) is broadband or monochromatic, it can still be used to control electronic transitions.

\subsubsection{Long-Wavelength Approximation}

To check the consistency of our result, we compute the transition rates of dipoles associated with  hydrogen transitions near a graphene layer ($\xi_q = 1$) and compare the results with the literature \cite{koppens2011graphene}. In the case of a z-dipole transition, $m_i=m_f=0$, so the only contribution to the transition rate is the $0$-th order vortex mode. As a result $\langle\phi_f|\hat{x}|\phi_i\rangle = \langle\phi_f|\hat{y}|\phi_i\rangle = 0$. Using the relation $\hat{p} = -\frac{i\hbar}{m} [\hat{x}, H_{ele}]$, we can show that $-i\omega\langle\phi_f|\hat{x}_j|\phi_i\rangle= \langle \phi_f|\hat{p}_j|\phi_i\rangle$, where $j$ labels the direction.

Considering a vortex mode whose wavelength is much larger than the atomic radius, the transition rate is only dependent on the field at the position of the atom.
\begin{align}
  \Gamma_{\text{z-dip}} &=  \frac{4\pi n\alpha\eta^3\omega}{\bar{\epsilon}_r}   \frac{\left|\bm{F}^j_{0,q}(0,0,z_0)\langle \phi_f|  \hat{p}_j | \phi_i\rangle \right|^2}{ m^2c^2} =  \frac{4\pi n\alpha\eta^3\omega}{\bar{\epsilon}_r} \frac{e^{-2q|z_0|}}{2} J_0(0)^2m^2\omega^2 |\langle \phi_f | \hat{z} | \phi_i\rangle|^2  \notag\\
  & =  \frac{2\pi n\alpha\eta^3\omega^3}{\bar{\epsilon}_r} e^{-2q|z_0|} |\langle \phi_f | \hat{z} | \phi_i\rangle|^2 .
\end{align}
The corresponding transition due to plane wave modes in vacuum is given by:
\begin{equation}
  \Gamma_{\text{free}} = \frac{4n\alpha \omega^3}{3c^2} |\langle \phi_f | \hat{z} | \phi_i\rangle|^2. \nonumber
\end{equation}
The presence of the polariton supporting graphene layers leads to an enhancement (Purcell factor) of:
\begin{equation}
  \frac{\Gamma_{\text{z-dip}}}{\Gamma_{\text{free}}} = \frac{3\pi}{2\bar{\epsilon}_r} \eta^3e^{-2q|z_0|}.
\end{equation}
Next we consider the decay rate for a dipole polarized in the $x$-direction. We perform this calculation in the Coulomb gauge because of its simplicity. Decay rates of course cannot depend on the choice of gauge \footnote{We do not need to change the unperturbed wavefunctions due to a theorem which states that transition amplitudes computed between states of the same energy are the same using the same unperturbed states in any gauge \cite{craig1984molecular}.}. Such a dipole can be seen as a superposition of an $m=1$ and an $m=-1$ dipole (i.e., a sum of oppositely oriented but circularly polarized dipoles). This means that the decay rate will have contributions from emission of an $m=1$ vortex and an $m=-1$ vortex. Since the dipole is $x$-polarized, only the $\cos\phi$ part of $e^{i\phi}$ or $e^{-i\phi}$ will contribute to the matrix elements. The total decay rate is therefore:
\begin{equation}
\Gamma_{x-dip}=2\times\frac{4\pi n\alpha\eta^3\omega^3}{\bar{\epsilon}_r}\Big|\int d\mathbf{r}\  \psi_g^*J_1(q\rho)Z_q(z)\cos\phi~ \psi_e  \Big|^2.
\end{equation}
Performing an expansion of the $J_1(q\rho)Z_q(z)$ term about the position of the atom $(0,0,z_0)$ and using the fact that $J_1(x) \approx x/2$ for small $x$, we have that the decay rate is:
\begin{equation}
\Gamma_{x-dip}=\frac{1}{2}\times\frac{4\pi n\alpha\eta^3\omega^3}{\bar{\epsilon}_r}|Z_q(z_0)|^2\Big|\int d\mathbf{r}\  \psi_g^*r\sin\theta\cos\phi ~\psi_e  \Big|^2.
\end{equation}
Setting $Z(z_0) = e^{-qz_0}$ and comparing the resulting expression to $\Gamma_{z-dip}$, the emission/absorption from an x-polarized dipole is half as much as from a $z$-polarized dipole with the same dipole moment. The same argument will show that for a $y$-polarized dipole, the emission is half as much as from a $z$-polarized dipole with the same dipole moment.  These decay rates for $z-$ and in-plane polarized dipoles match the results obtained in \cite{koppens2011graphene} for the decay rates of emitters into graphene plasmons.

\subsection{Second-Order Processes}

\subsubsection{Spontaneous Emission in the Angular Momentum Basis}

In \cite{Rivera2016}, two-polariton spontaneous emission rates for $s\rightarrow s$ transitions are computed. We now establish the formalism for the spontaneous emission of two vortices, thus opening up a path for analyzing the quantum optics of polaritons beyond first-order processes.  The differential spontaneous emission into vortices at frequencies $\omega_{q}$ and $\omega_{q'}$ with wavevectors $q$ and $q'$ and  angular momenta $m$ and $m'$ is \footnote{Note that strictly speaking, we are ignoring the effect of intermediate levels in between the initial and final states. We can assume there are no such states, or realize that their effect is to create a radiative cascade which is effectively first-order in perturbation theory. To optimize two-photon spontaneous emission, one would want to supress or avoid radiative cascades as much as possible. It is described how to do so in \cite{Rivera2016}. For absorption, one simply wants to avoid exciting at frequencies that would resonantly excite a state in-between the initial and final states.}:
\begin{equation}
\Gamma = \frac{2\pi}{\hbar^2}\left(\frac{1}{2}\frac{L^2}{\pi^2}\int dq\int dq' \sum_{m,m'} \right) \Big| \sum_{i_1} \frac{e\phi_{g,i_1}e\phi_{i_1,e}}{E_e-E_{i_1}+i0^+} \Big|^2\delta(\omega_0 - \omega_{q}-\omega_{q'}).
\end{equation}
which inserting our potential operator (\req{21}) yields:
\begin{align}
\frac{d\Gamma}{d\omega}&\left(e,0 \rightarrow g,qm,q'm'  \right) = \frac{1}{4}\pi\alpha^2\frac{c^2}{v_g(\omega)v_g(\omega_0-\omega)}\frac{ \omega(\omega_0-\omega)}{\bar{\epsilon}^2_r\xi_{q}(\omega)\xi_{q'}(\omega_0-\omega')}\times \nonumber \\
\Big|\sum_{n}&\frac{\langle g,qm,q'm' |J_{m'}e^{-im'\phi}(q'\rho)Z_{q'}(z)|n,qm\rangle \langle n,qm|J_{m}(q\rho)e^{-im\phi}Z_{q}(z)|e,0\rangle }{\omega_e - \omega_n -\omega_{q} } \nonumber \\ +&\frac{\langle g,qm,q'm' |J_{m}(q\rho)e^{-im\phi}Z_{q}(z)|n,q'm'\rangle \langle n,q'm'|J_{m'}(q'\rho)e^{-im'\phi}Z_{q'}(z)|e,0\rangle }{\omega_e - \omega_n -\omega_{q'}}  \Big|^2.
\end{align}
For an $s\rightarrow s$ transition in the long-wavelength approximation, $\delta\ell = 1$ for each virtual transition, meaning that the virtual states can only have $m=0$ or $m=\pm 1$. The total differential decay rate is thus
\begin{align}
& \frac{d\Gamma}{d\omega}(e,0\rightarrow g,qq')\equiv\sum_{m}\frac{d\Gamma}{d\omega} (e,0 \rightarrow g,qm,q'(-m)) =  \nonumber \\
&\frac{d\Gamma}{d\omega} (e,0 \rightarrow g,q0,q'0) +\frac{d\Gamma}{d\omega} (e,0 \rightarrow g,q1,q'(-1))+\frac{d\Gamma}{d\omega} (e,0 \rightarrow g,q(-1),q'1) = \nonumber \\
&\frac{d\Gamma}{d\omega} (e,0 \rightarrow g,q0,q'0) + 2\frac{d\Gamma}{d\omega} (e,0 \rightarrow g,q1,q'-1).
\end{align}
Using the fact that the matrix elements for the $|m|=1$ virtual transitions is $\sqrt{2}$ smaller than that for the $m=0$ virtual transitions (see end of Sec. V.A.1), the total decay rate is:
\begin{equation}
\frac{d\Gamma}{d\omega}(e,0\rightarrow g,qq')\equiv\sum_{m}\frac{d\Gamma}{d\omega} (e,0 \rightarrow g,qm,q'(-m)) = \frac{3}{2}\frac{d\Gamma}{d\omega} (e,0 \rightarrow g,q0,q'0)
\end{equation}
Within the dipole approximation, the main contribution to the matrix elements in the sum over intermediate states comes from the $zJ_0(0)\frac{dZ}{dz}\Big|_{z=z_0} = zZ'(z_0)$ term of the Taylor expansion of the field mode. Taking the emitter to be in the air region with $Z = e^{-qz_0}$, and defining $\eta(\omega) \equiv \frac{q(\omega)c}{\omega}$, the differential decay rate is:
\begin{align}
&\frac{3\pi\alpha^2}{8c^4\bar{\epsilon}_r^2}\left(\frac{1}{\xi_{q}(\omega)}\frac{c}{v_g(\omega)}\eta^2(\omega)e^{-2\eta(\omega)k(\omega)z_0}\right)\left(\frac{1}{\xi_{q'}(\omega_0-\omega)}\frac{c}{v_g(\omega_0-\omega)}\eta^2(\omega_0-\omega)e^{-2\eta(\omega_0-\omega)k(\omega_0-\omega)z_0}\right) \times \nonumber \\
&\omega^3(\omega_0-\omega)^3 \Big|z_{gn}z_{ne}\left( \frac{1}{\omega_e - \omega_n + \omega-\omega_0 } + \frac{1}{\omega_e - \omega_n -\omega} \right)  \Big|^2,
\end{align}
where $z_{ab} \equiv \langle a|z|b \rangle$. The two-photon differential emission rate in free-space, by comparison is \cite{breit1940metastability}
$$
\frac{d\Gamma}{d\omega}\Big|_{free~space} = \frac{4}{3\pi c^4}\alpha^2\omega^3(\omega_0-\omega)^3\Big| \sum_n z_{gn}z_{ne}\left( \frac{1}{\omega_e - \omega_n + \omega -\omega_0} + \frac{1}{\omega_e - \omega_n - \omega} \right)  \Big|^2. 
$$
\begin{align}
\frac{d\Gamma/d\omega\Big|_{polaritons}}{d\Gamma/d\omega\Big|_{free~space}} =& \frac{9\pi^2}{32\bar{\epsilon}_r^2}\left(\frac{1}{\xi_{q}(\omega)}\frac{c}{v_g(\omega)}\eta^2(\omega)e^{-2\eta(\omega)k(\omega)z_0}\right) \nonumber \\ &\left(\frac{1}{\xi_{q}(\omega_0-\omega)}\frac{c}{v_g(\omega_0-\omega)}\eta^2(\omega_0-\omega)e^{-2\eta(\omega_0-\omega)k(\omega_0-\omega)z_0}\right).
\end{align}
This result applies in the lossless limit of any highly confined polaritonic mode. All one needs is their $\xi_{q}$ factors and their group velocities, which will be of order $c/\eta$. In order to check the validity of our result, we take the special case of a 2D plasmonic material. For graphene in the Drude model, $c/v_g = 2\eta$ and $\xi_{q} = 1$, meaning that the two-photon enhancement in the angular momentum basis is $\frac{9\pi^2}{8\bar{\epsilon}^2_r}\eta^3(\omega)\eta^3(\omega_0-\omega)\exp\left[-2\frac{z_0}{c}(\omega\eta(\omega)+(\omega_0-\omega)\eta(\omega_0-\omega))\right]$, in agreement with the result in \cite{Rivera2016}.

\subsubsection{Absorption of Two-Vortices}
The calculation of absorption of two vortices will be quite parallel to the calculation for emission except that factors of photon occupation number will be present. Therefore, in the lossless limit, the differential rate of two-vortex absorption is (assuming only $m=0$ vortices are present):
\begin{equation}
\frac{d\Gamma}{d\omega}(e,0\rightarrow g,q,q')= \frac{d\Gamma}{d\omega} (e,0 \rightarrow g,q0,q'0) n_0(\omega)n_0(\omega_0-\omega),
\end{equation}
where $n_0$ is the number of photons with zero angular momentum. The photon number is calculated in the same manner as detailed in the section on first-order transitions. The selection rule is of course that $\Delta m = 0$. More generally, the selection rule for an absorption transition involving two vortices of angular momenta $m_1$ and $m_2$ is that $\Delta m = m_1+m_2$. 

\bibliography{Arxiv.bib}